# Ferrotoroidic Ground State in a Heterometallic $Cr^{III}Dy^{III}_6$ Complex Displaying Slow Magnetic Relaxation

Kuduva R. Vignesh,[1] Alessandro Soncini,[*,2] Stuart K. Langley,[3] Wolfgang Wernsdorfer,[4] Keith S. Murray,[*,5] and Gopalan Rajaraman[*,6]

Toroidal quantum states are most promising for building quantum computing and information storage devices as they are insensitive to homogeneous magnetic fields, but interact with charge and spin currents, allowing this moment to be manipulated purely by electrical means. Coupling molecular toroids into larger toroidal moments via ferrotoroidic interactions can be pivotal not only to enhance ground state toroidicity, but also to develop materials displaying ferrotoroidic ordered phases, which sustain linear magneto-electric coupling and multiferroic behaviour. However, engineering ferrotoroidic coupling is known to be a challenging task. Here we have isolated a $\{Cr^{III}Dy^{III}_6\}$ complex, which exhibits the much sought-after ferrotoroidic ground state with an enhanced toroidal moment, solely arising from intramolecular dipolar interactions. Moreover, a theoretical analysis of the observed sub-Kelvin zero-field hysteretic spin dynamics of $\{Cr^{III}Dy^{III}_6\}$ reveals the pivotal role played by ferrotoroidic states in slowing down the magnetic relaxation, in spite of large calculated single-ion quantum tunnelling rates.

[1]IITB-Monash Research Academy, IIT Bombay, Mumbai-400076, India. [2]School of Chemistry, University of Melbourne, Victoria-3010, Australia. [3]School of Science and the Environment, Chemistry Division, Manchester Metropolitan University, Manchester, UK. [4]Institute Neel, CNRS, F-38000 Grenoble, France and Institute of Nanotechnology, Karlsruhe Institute of Technology, 76344 Eggenstein-Leopoldshafen, Germany. [5]School of Chemistry, Monash University, Victoria-3800, Australia. [6]Department of Chemistry, Indian Institute of Technology Bombay, Mumbai-400076, India. Correspondence and requests for materials should be addressed to A.S., K.S.M., G.R. (email: asoncini@unimelb.edu.au; keith.murray@monash.edu; rajaraman@chem.iitb.ac.in).

The magnetic behaviour of molecular coordination complexes continues to intrigue scientists around the world, revealing many interesting physical properties and offering many potential applications such as new storage and information processing technologies.[1-3] Fundamental research into, for example, single-molecule magnets (SMMs),[4-6] spin-crossover[7] and magnetic systems with toroidal moments[8-19] are recognized as important areas of molecular magnetism. SMMs exhibit slow relaxation of the magnetization, acting as nano-magnets below their blocking temperature.[2,4-6] Molecular coordination complexes with a toroidal arrangement of local magnetic moments are rare, but are of great interest as they have several potential applications such as quantum computation, molecular spintronics devices, and the development of magneto-electric coupling for multiferroic materials.[20-22] Toroidal moments at a molecular level were first predicted[12] and observed[9,17] in 2008, in strongly anisotropic metal complexes with a ring topology, having in–plane magnetic axes tangential to the ring, and weak nearest neighbour exchange coupling of the appropriate sign. In particular, the observation of a toroidal texture[9,17] in the ground state of the $\{Dy^{III}_3\}$ triangular system,[8] generated great interest in this area, with state-of-the-art theoretical[9,14] and experimental[10,17,18] techniques being employed to probe toroidal states of this prototype and subsequent related molecules. Other than triangular $\{Dy_3\}$ complexes the toroidal moments have been reported as well for rhombus $\{Dy_4\}$[23-25] and $\{Dy_6\}$ wheel[26] complexes.

If two molecular rings exhibiting toroidal states are connected, for example via a 3d ion, then coupling between the two toroidal moments leading to an enhancement of the collective toroidal moment may occur, which is a prerequisite to achieve a molecular ferrotoroidically ordered phase and the development of molecule-based multiferroics.[11,13,18,27] In order to isolate materials with the above mentioned properties we target coordination complexes containing anisotropic 4f ions. In contrast to the great deal of work on the synthesis of 3d-4f coordination complexes using 3d ions such as $Mn^{III}$, $Fe^{III}$ and $Co^{II}$,[28-32] there have been few reports of studies, both structurally and magnetically, on mixed Cr(III)-Ln(III) systems. We have recently shown, however, that the combination of the isotropic $Cr^{III}$ ion and the anisotropic $Dy^{III}$ ion resulted in a family of SMMs with long relaxation times, relative to other lanthanide-based SMMs.[33-36] With this in mind, we have chosen to expand our studies, utilizing chromium(III) nitrate, with various lanthanide(III) ions, with carboxylic acid pro-ligands.

Herein, we describe the synthesis, structural characterization, and magnetic properties of a heterometallic complex of formula [$Cr^{III}Dy^{III}_6(OH)_8(ortho\text{-tol})_{12}(NO_3)(MeOH)_5$]·3MeOH (**1**), where *ortho*-tol = ortho-toluate. Complex **1** displays slow magnetic relaxation and SMM behaviour at temperatures below 2 K. Furthermore, in {$Cr^{III}Dy^{III}_6$} we find, for the first time, a ferrotoroidically coupled ground state fully determined by dipolar coupling between the two con-rotating toroidal triangles (see Supplementary Information for a detailed comparison of our findings with previous studies of coupled molecular {$Dy^{III}_3$} toroids).[10,11,18] The ferrotoroidically coupled ground state thus leads to an enhanced toroidal moment in the ground state for the {$Cr^{III}Dy^{III}_6$} complex, which is shown to play a central role in the observed magnetisation dynamics featuring zero-field hysteresis.

## Results

**Synthesis and Magnetic properties**. Compound **1** was synthesized by the reaction of $Cr(NO_3)_3·9H_2O$ and $Dy(NO_3)_3·6H_2O$, with *ortho*-toluic acid in acetonitrile at ambient temperature. The solvent was then removed and re-dissolved in MeOH/$^i$PrOH (see Supplementary method). Suitable single crystals (pale purple color) for X-ray analysis were isolated after allowing the solvent to slowly evaporate.

Single crystal X-ray analysis reveals that compound **1** crystallizes in the triclinic space group, *P*-1 (see Supplementary Table 1 for full crystallographic details). The asymmetric unit contains half the complex, (three $Dy^{III}$ ions and one half of the $Cr^{III}$ ion) which lies upon an inversion centre. Compound **1** is a heterometallic heptanuclear complex consisting of a single $Cr^{III}$ ion and six $Dy^{III}$ ions (Figure 1). The low $Cr^{III}$ to $Dy^{III}$ ratio of 1:6 in **1** is likely a consequence of the limited solubility of $Cr(NO_3)_3·9H_2O$ in MeCN. The metallic core is based on two triangular $Dy^{III}$ units which lie above and below a single central $Cr^{III}$ ion, revealing two vertex sharing trigonal pyramids or tetrahedra. The metallic core is stabilized by eight $\mu_3$-hydroxide, twelve *ortho*-toluate, with MeOH and [$NO_3$]$^-$ ligands. Six of the $\mu_3$ hydroxide ligands bridge two $Dy^{III}$ ions to the central $Cr^{III}$ ion, while the remaining two bridge the three $Dy^{III}$ ions that make up each triangle. Six of the *ortho*-toluate ligands each bridge a $Dy^{III}$-$Dy^{III}$ triangular edge, while six are found to chelate, each to a single $Dy^{III}$ ion. Terminal MeOH ligands coordinate to all six $Dy^{III}$ ions. It is found, however, that, at two of the $Dy^{III}$ sites, disordered

MeOH and nitrate ions (50:50 occupancy) are present. The $Cr^{III}$ ion is six coordinate with an octahedral geometry, while the six $Dy^{III}$ ions are eight coordinate.

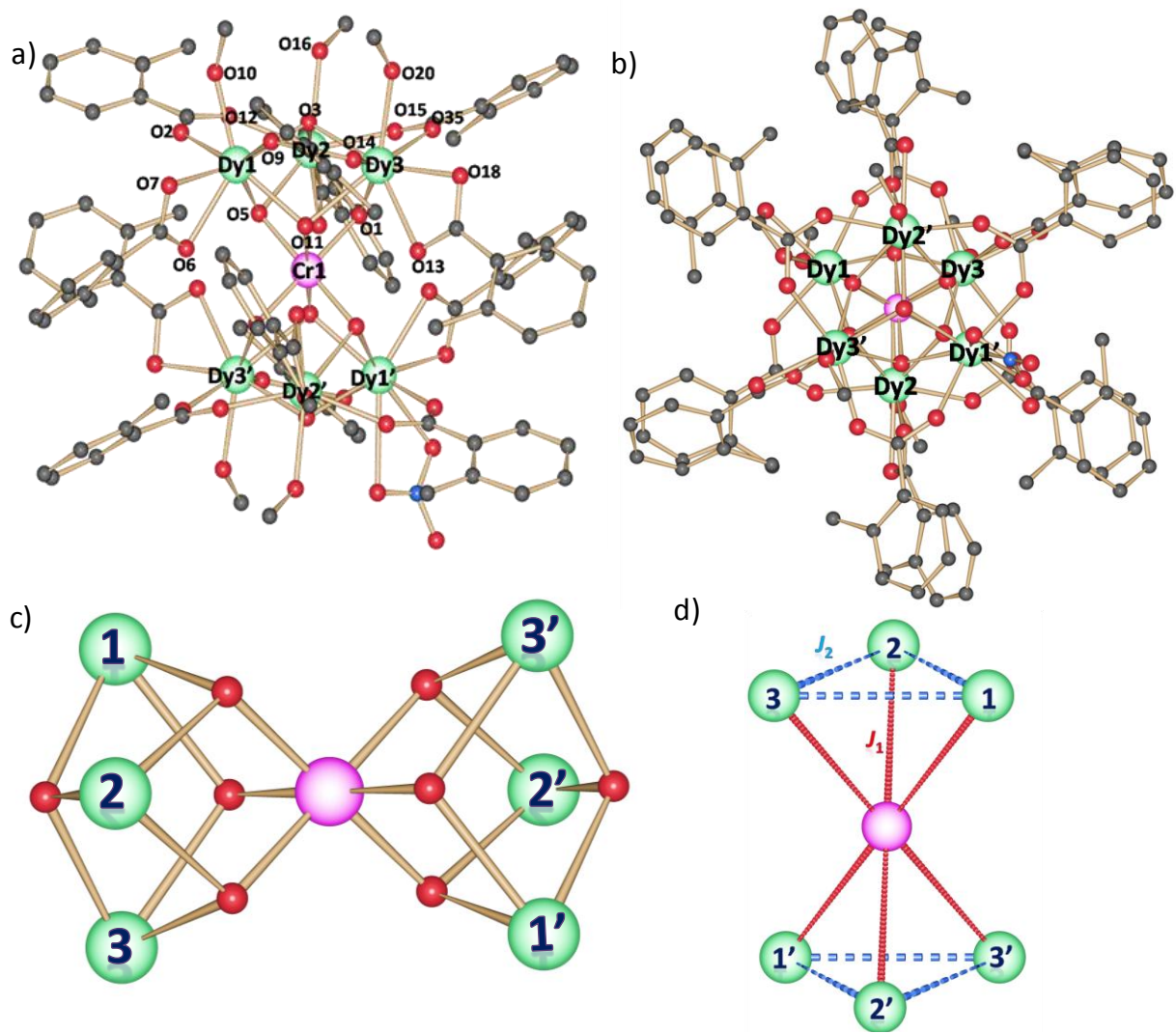

**Figure 1 | Molecular structure and exchange pathways.** (a) The molecular structure of complex **1**. The solvent and H atoms are omitted for clarity. Color scheme; $Cr^{III}$, pink; $Dy^{III}$, green; O, red; N, blue; C, light grey; (b) Top view of the molecular structure of **1**; (c) Metal topology found in **1** with (d) magnetic exchange pathways $J_1$, and $J_2$ highlighted.

We have examined the structural distortions at individual $Dy^{III}$ sites using SHAPE software.[37,38] The geometry of each $Dy^{III}$ ion is best described by a triangular dodecahedron. The deviation of 2.7 for Dy1 and Dy1', 1.2 for Dy2 and Dy2', 1.5 for Dy3 and Dy3' are observed with respect to the ideal triangular dodecahedron. Selected bond lengths and $Dy^{III}$-O-$Dy^{III}$ and $Cr^{III}$-O-$Dy^{III}$ bond angles are given in Supplementary Table 2. The $Dy^{III}$-O bond lengths are in the range, 2.391-2.492 Å. In the {$Dy^{III}_3$} triangular unit, the bond distance between Dy1-Dy2, Dy1-Dy3 and Dy2-Dy3 is found to be 3.749 Å, 3.767 Å and 3.780 Å, respectively and the Dy3-Dy1-Dy2, Dy1-Dy2-Dy3 and Dy2-Dy3-Dy1 bond angles are 55.99°, 60.52° and 59.49°, respectively. The average Dy-($\mu_3$-OH)-Dy angle is 106.0°, while the average Dy-($\mu_3$-OH)-Dy angle also bridging to the $Cr^{III}$ ion is slightly smaller at 103.4°. The centroid to centroid distance between the two triangular units is found to be 5.38 Å. A shorter distance, compared to other linked {$Dy^{III}_3$} triangles[11] (5.64 Å) is likely to yield stronger dipolar coupling between the two {$Dy^{III}_3$} units. Packing diagrams of **1**, viewed along the *a*, *b* and *c* axes and of a neighboring pair are shown in Supplementary Figure 1. There are H-bonds linking adjacent {$Cr^{III}Dy^{III}_6$} complexes via MeOH⋯MeOH(solv)⋯O(carb) groups, combined with edge to face π- π interactions.

Dc magnetic susceptibility data were collected for **1** and the variation of $\chi_M T$ with temperature is shown in Figure 2a. The room temperature $\chi_M T$ product of 87.16 cm$^3$ K mol$^{-1}$ is in agreement with the value expected (86.9 cm$^3$ K mol$^{-1}$) for one $Cr^{III}$ (S=3/2, *g* = 2, C= 1.875 cm$^3$ K mol$^{-1}$) and six $Dy^{III}$ (S=5/2, L=5, $^6H_{15/2}$, *g* = 4/3, C= 14.17 cm$^3$ K mol$^{-1}$) ions that are non-interacting. As the temperature is reduced the $\chi_M T$ product (*H* = 1 T) decreases gradually between room temperature and 50 K, before a more rapid decrease below this temperature, reaching a value of 14.37 cm$^3$ K mol$^{-1}$ at 1.8 K. The decrease in $\chi_M T$ at higher temperatures is attributed to the depopulation of the excited $m_J$ Stark states of the $Dy^{III}$ ions, while the more rapid decrease at low temperatures is indicative of the presence of dominant antiferromagnetic exchange interactions. The low temperature $\chi_M T$ value of 14.37 cm$^3$ K mol$^{-1}$ at 2K is higher than that expected for a single paramagnetic Cr(III) ion suggest that there are several close lying excited states including that of $Dy^{III}$ ion which possess significant magnetic moment.

The isothermal *M vs H* plots (Figure 2b, and Supplementary Figure 2) at low fields reveal a non-linear, S-shaped curve at 2 K and 3 K, which suggests the presence of a toroidal moment[8-14] and/or possible blockage of the magnetization vector and therefore slow magnetic relaxation. Above 4 K, the plots display a rapid increase in magnetization up to ~ 2 T followed by a steady increase and almost saturating at 5T (Figure 2b). The simulation of the plots in Fig. 2 are discussed later.

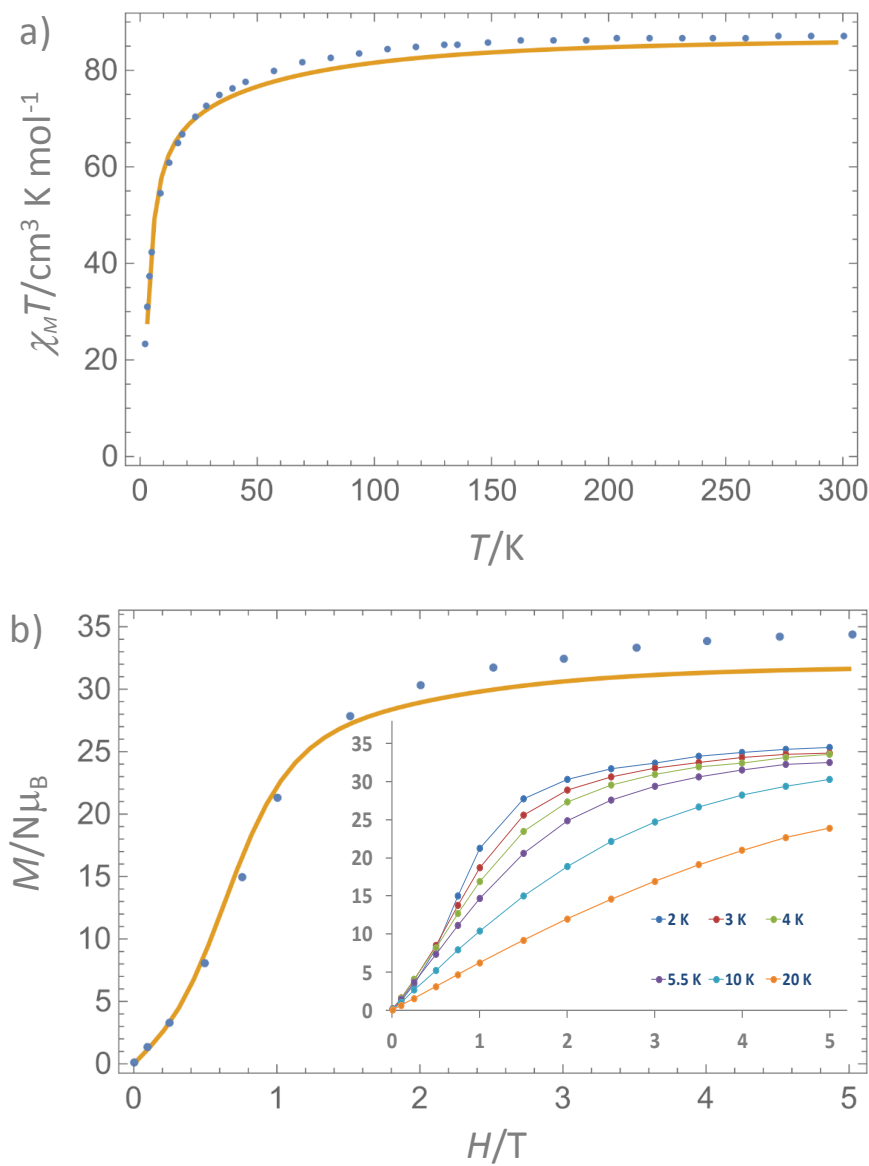

**Figure 2 | Susceptibility and magnetization plots.** The measured (blue circles) and simulated (via the *ab initio*-parameterized model, orange solid line) plot of (a) Magnetic susceptibility versus temperature at 1T and (b) *M* vs *H* isotherms at 2K for complex **1**. (inset) *M* versus *H* isotherms for complex **1** at 2, 3, 4, 5.5, 10 and 20 K (solid lines just join the points here).

To probe for SMM behavior, the magnetization dynamics were investigated via alternating current (ac) susceptibility measurements as a function of both temperature and frequency. A 3.5 Oe ac field was employed, utilizing both a zero and a 2000 Oe static dc field. A non-zero out-of-phase magnetic susceptibility component ($\chi''$) is observed, at $H_{dc}$= 0 Oe, however, no maxima are found upon reducing the temperature down to 1.8 K (Supplementary Fig. 3). This is also the case for when $H_{dc}$ = 2000 Oe (Supplementary Fig. 3). This does not prove SMM behavior, but suggests the possibility of such, with a small energy barrier to magnetic reorientation and fast relaxation times, even at 1.8 K. As a consequence of the fast magnetic relaxation times, even at temperatures below 2 K, it is suggested that the low field magnetization behavior points to the presence of a toroidal magnetic moment.

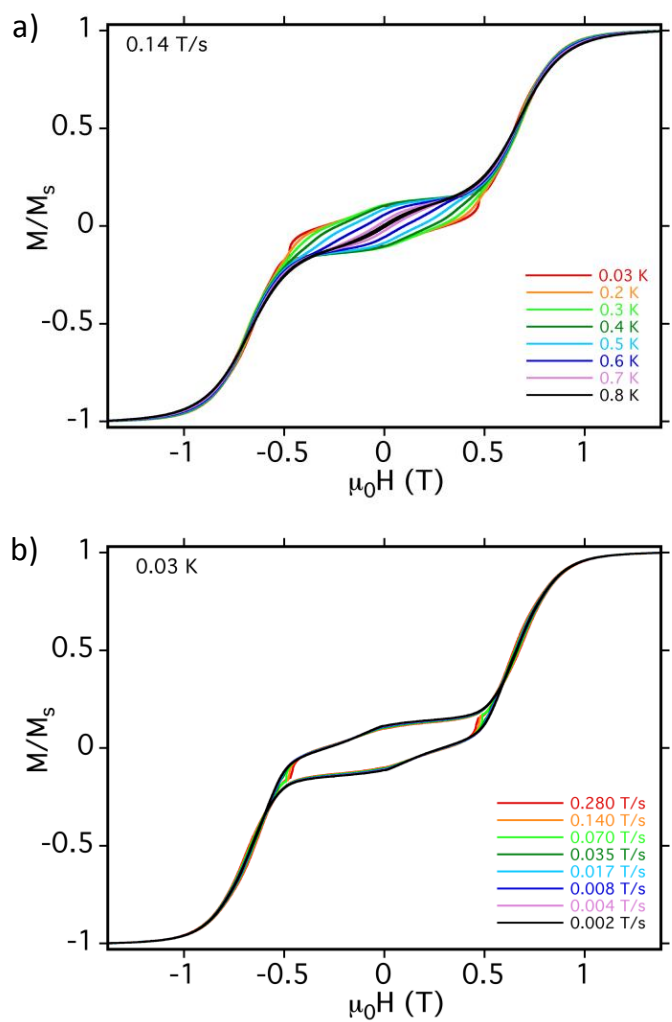

**Figure 3 | Single-crystal studies.** Single-crystal magnetization (M) vs. applied field measurements (μ-SQUID) for complex **1** at (a) 0.03 K to 0.8 K with the scan rate of 0.14 Ts$^{-1}$; (b) with different field sweep rates at 0.03 K.

Single crystals of **1** were studied using the micro-SQUID technique at various temperatures and sweep rates for two different orientations of the molecule.[39] The curve displays a stepped shape of the magnetization (Figure 3 and Supplementary Fig. 4), similar to that observed by the archetypal triangular toroidal system.[8] The position of the step ($H_s$) depends on the orientation of the magnetic field with respect to the plane of the triangles, as expected on the basis of the in-plane anisotropy predicted by our model (*vide infra*). Thus, when the field is parallel to two inversion-related tangential Dy magnetic axes, and perpendicular to none, this leads to an in-plane easy axis, while a magnetic field perpendicular to that direction, thus perpendicular to two Dy easy axes and parallel to none, leads to an in-plane hard axis. In particular, when the field is applied between the $\{Dy^{III}_3\}$ triangles (i.e. along the Dy-Dy bond vector, Figure 3a) hysteresis is observed below 0.8 K, with the coercive field widening on cooling ($H_c$= 0.6 T at 0.03 K, with a sweep rate of 0.28 T/s). This behavior is characteristic of an SMM, with slow zero-field relaxation. We then observe a large step in the magnetization at about $H_s$ = 0.7 T. Application of the magnetic field perpendicular to the Dy-Dy bond vector results in a reduction of coercivity ($H_c$= 0.5 T at 0.03 K, with a sweep rate of 0.28 T/s, Supplementary Fig. 4). Upon comparison of the hysteresis profile of complex **1** with the archetypal $\{Dy_3\}$ toroidal complex,[8] we find that profile for **1** appears to be superior comparing the coercitivity. This suggests that coupling between the two $\{Dy^{III}_3\}$ triangles in **1** enhances the zero-field slow relaxation properties of the system. A theoretical model explaining this behaviour is developed later in the paper.

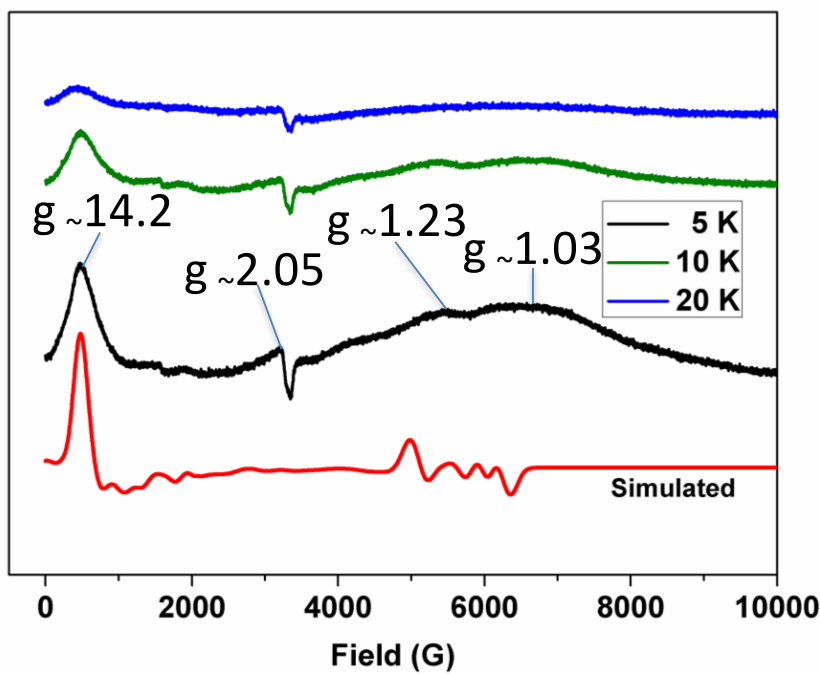

**Figure 4. EPR spectroscopy.** Powder EPR spectra of **1** at X-band frequency at 5 K, 10 K and 20 K and the simulated curve.

We have also recorded EPR spectra at the X-band frequency at 5, 10 and 20 K (See Figure 4). The EPR spectrum recorded at 5 K reveals distinct features at very large *g*-values ($g$~14.2). When we increased the temperature we found that the intensity of this signal decreases. There are also weak features at $g$~2.05, $g$~1.2 and $g$~1.03 and the intensities of these features also decrease upon increasing the temperature. To gain an understanding on the nature of these EPR spectra, we have simulated the EPR by means of the XSOPHE simulation suite,[40,41] using a $\{Cr^{III}Dy^{III}_3\}$ model employing a pseudo $S$=1/2 state for each $Dy^{III}$ ion and a $S$=3/2 state for the $Cr^{III}$ ion. *Ab initio* computed *g*-anisotropies, directions and *J* values are given as inputs (See below for details) along with the Dy⋯Dy and Dy⋯Cr distances from the X-ray structure. Besides these values, the orientation of the g-anisotropies computed from *ab intio* calculations were also taken into account. A small perturbation to the Euler angles without altering any other parameters yield reasonable fit to the experimental spectrum recorded at 5 K (See Figure 4; The further details of simulation are given in Supplementary Note 1), offering confidence on the estimated parameters. However, the lines appearing at g~1.23 and g~1.03 are much broader than than they appear in the simulation and this may be attributed to a fact that only $\{Cr^{III}Dy^{III}_3\}$ has been employed in the simulation and not the full $\{Cr^{III}Dy^{III}_6\}$ Hamiltonian. Multi-frequency EPR including HF-EPR spectra are required, in future, to independently obtain the spin Hamiltonian parameters.[42]

**Theoretical Analysis: Characterization of a Ferrotoroidic Ground State**

To explain the experimental observations, we performed *ab-initio* calculations of electronic structure and magnetic properties, using MOLCAS 7.8[43], on individual $Dy^{III}$ and $Cr^{III}$ centers. The computed orientation of the anisotropy axes is shown in Figure 5. In particular, we employed the *ab initio* $M_J$-decomposition of the single-ion thermally-isolated ground Kramer's doublet wavefunctions along the *ab initio g*-tensor principal axis, to set up a model Hamiltonian for intramolecular magnetic coupling including dipolar coupling between all pairs of ions, which is parameters-free, intra-ring $Dy^{III}$-$Dy^{III}$ superexchange interactions parameterized by a single coupling constant $J_2$, and $Dy^{III}$-$Cr^{III}$ superexchange interactions parameterized by a single coupling constant $J_1$. The coupling parameters $J_1$ and $J_2$ were evaluated via DFT calculations. The well-known dipolar Hamiltonian reads:

$$H_{\text{dip}} = \frac{\mu_0}{4\pi} \sum_{p,q} \left( \frac{\boldsymbol{M}_p \cdot \boldsymbol{M}_q}{|\boldsymbol{R}_{pq}|^3} - 3 \frac{(\boldsymbol{M}_p \cdot \boldsymbol{R}_{pq})(\boldsymbol{M}_q \cdot \boldsymbol{R}_{pq})}{|\boldsymbol{R}_{pq}|^5} \right)$$

(1)

where $\boldsymbol{M}_p$ is the magnetic moment of the $p^{\text{th}}$ ion, and $R_{pq}$ the distance between ions p and q. The superexchange contribution is modeled by an isotropic Heisenberg Hamiltonian[44]:

$$H_{\text{exch}} = -J_2 \sum_{p,q} (\boldsymbol{S}_p \cdot \boldsymbol{S}_q + \boldsymbol{S}_{p'} \cdot \boldsymbol{S}_{q'}) - J_1 \sum_q (\boldsymbol{S}_q + \boldsymbol{S}_{q'}) \cdot \boldsymbol{S}_{Cr}$$

(2)

where $S_q$ ($S_{\text{Cr}}$) are the true spin moments of the Dy (Cr) ions, with the primed and unprimed subscripts labelling Dy ions belonging to different triangles. We note here that when the simple isotropic exchange Hamiltonian is projected on the thermally isolated ground KDs (Kramers doublet) it becomes a strongly anisotropic non-collinear Ising Hamiltonian, a widely employed protocol for {3d-4f} systems.

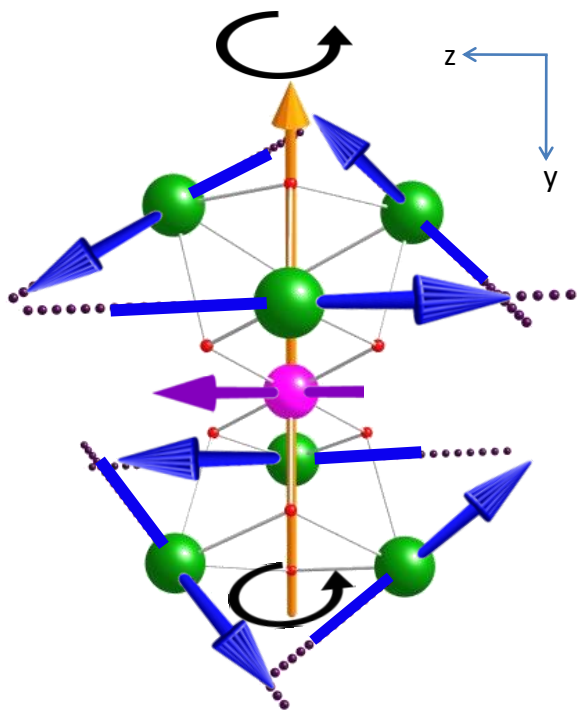

**Figure 5 | Orientation of the magnetic anisotropy axes.** The directions of the local anisotropy axes in the ground Kramers doublet on each Dy site (dotted lines) in **1**. Blue arrows are the local magnetic moment in the ground exchange doublet. Black arrows show the con rotation of the toroidal magnetic moment and the yellow arrow is the $S_6$ symmetry axis.

To estimate the low-energy wavefunctions and magnetic anisotropy for each of the seven ions in **1**, we have undertaken CASSCF+RASSI-SO calculations on the individual $Dy^{III}$ centers.[43] The calculations yielded the following *g*-tensor principal values: [Dy1; $g_x$ = 0.0523, $g_y$ = 0.0927 and $g_z$ = 19.5707]; [Dy2; $g_x$ = 0.0737, $g_y$ = 0.0979 and $g_z$ = 19.4723]; [Dy3; $g_x$ = 0.0233, $g_y$ = 0.0361 and $g_z$ =19.6059] and [Cr; $g_x$ = $g_y$ = $g_z$ = 2.002] (see Supplementary Tables 3 and 6). The symmetry related Dy1', Dy2' and Dy3' ions possess essentially the same *g*-tensor. Although the two triangles are equivalent, Dy1 and Dy1' slightly differs due to coordination of methanol in Dy1 and nitrate in Dy1'. These data, along with the *J* value are found to yield a good fit to the experimental susceptibility data (see Figure 2, *vide supra*). A qualitative mechanism developed based on single-ion Dy(III) is discussed in detail in the ESI (See Supplementary Figure 6 and Supplementary Note 2).

The core structure of the molecule has a pseudo $S_6$ axis passing through the $Cr^{III}$ ion and the center of both of the $\{Dy^{III}_3\}$ triangular units (Figure 5). The local principal anisotropy axes are found from the calculations to lie in the $\{Dy^{III}_3\}$ plane with an out-of-plane angle of 0.29, 4.5 and 4.7° for Dy1, Dy2 and Dy3, respectively. The $Dy^{III}$ magnetic axes are also found to be almost perfectly aligned with the tangents to an ideal circumference enclosing the triangules (the angle of the anisotropy axis with these tangential directions are in the range of 1.1 to 7.9°). The computed energies of the eight low-lying Kramer's doublets (KDs) reflect that there are three types of $Dy^{III}$ ions in the complex (see Supplementary Tables 4 and 5), although the ground states of all $Dy^{III}$ ions consist of almost pure atomic $|\pm 15/2\rangle$ KDs. The energy gap between the ground and the first excited state KDs are found to be 142.8 cm$^{-1}$, 121.9 cm$^{-1}$ and 152.7 cm$^{-1}$ for Dy1, Dy2 and Dy3, respectively. For the $Cr^{III}$ ion calculations yield isotropic *g*-tensors (see Supplementary Table 3) and an axial zero-field splitting of $\sim 1.0 \times 10^{-4}$ cm$^{-1}$.

Thus the *ab initio* calculations suggest that magnetic coupling can be well described by projecting Hamiltonians Equation (1) and (2) on the basis of the ground KDs (Dy) and quartet (Cr) only, leading to a 256-dimensional product space with basis $|\boldsymbol{m}, M_{Cr}\rangle$, where $\boldsymbol{m}$ = [$m_1$, $m_2$, $m_3$, $m_{1'}$, $m_{2'}$, $m_{3'}$] with $m_q$= ±1. We further assume that such KDs are pure $|\pm 15/2\rangle$ atomic states quantized along the local anisotropy axis, which is assumed to be exactly in the triangle's plane and along the tangential direction, so that $\hat{S}_{t,q}|m_q\rangle = m_q(5/2)|m_q\rangle$, and $\hat{J}_{t,q}|m_q\rangle = m_q(15/2)|m_q\rangle$ ($\hat{S}_{t,q}$ and $\hat{J}_{t,q}$ are the spin and total angular momenta operators along the tangential direction for the $q^{th}$ $Dy^{III}$ ion), and a seventh quantum number $M_{Cr} = \pm 3/2, \pm 1/2$ labeling the spin state for the Cr ion. Finally, given the quasi $S_6$ symmetry, in our model we assume two equilateral triangles with radius $r = 2.17$ Å, their planes being at distance $h = 5.38$ Å (see Figure 6, $r$ and $h$ are average experimental values). Thus a ferrotoroidic (FT) state (con-rotating toroidal moments $\pm \tau_1$, $\pm \tau_2$ on the two triangles) corresponds to $|\pm 1, \pm 1, \pm 1, \pm 1, \pm 1, \pm 1, M_{Cr}\rangle \equiv |\pm \tau_1, \pm \tau_2, M_{Cr}\rangle$, while antiferrotoroidic (AFT) states (counter-rotating toroidal moments) corresponds to $|\pm 1, \pm 1, \pm 1, \mp 1, \mp 1, \mp 1, M_{Cr}\rangle \equiv |\pm \tau_1, \mp \tau_2, M_{Cr}\rangle$. Due to the large value of the ground KDs angular momenta-projections in $|\boldsymbol{m}, M_{Cr}\rangle$, Hamiltonians Equation (1) and (2) are both diagonal on such basis, and $|\boldsymbol{m}, M_{Cr}\rangle$ represent the low-energy exchange-coupled states of **1**. The corresponding energies can be written as a sum of a superexchange contribution, an intra-ring dipolar contribution, an inter-ring dipolar contribution, and $Dy^{III}$-$Cr^{III}$ dipolar contribution.

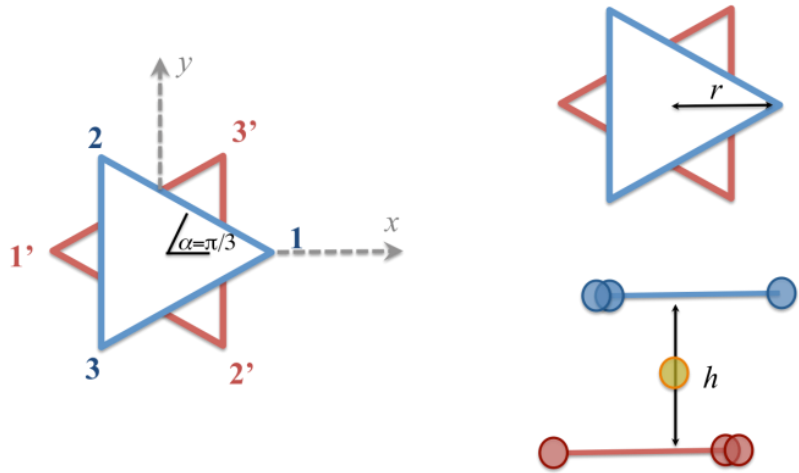

**Figure 6 | Schematic diagram to describe exchange coupling.** Schematic representation of the idealised geometry used to describe magnetic coupling in {Dy$_6$Cr} via Hamiltonians Equation (1) and (2).

For the *exchange energies* we get (sum over $q$ is understood modulus 3, so that $m_{3+1} = m_1$):

$$E_{\text{exch},\boldsymbol{m}} = \frac{25}{8} J_2 \sum_{q=1}^{3} \left( m_q m_{q+1} + m_{q'} m_{q'+1} \right) - J_1 M_{Cr} \mathcal{M}_{\boldsymbol{m}}$$

(3)

where, defining the angular coordinates of the six $Dy^{III}$ ions as in Figure 6 ($\alpha_1=0$, $\alpha_2=2\pi/3$, $\alpha_3=4\pi/3$, $\alpha_{1'} = \pi$, $\alpha_{2'} = -\pi/3$, $\alpha_{3'} = \pi/3$), the total spin projection ($\mathcal{M}_m$) of the six $Dy^{III}$ ions in $|m, M_{Cr}\rangle$ is given by Equation (4) (sum is over all $Dy^{III}$ centers):

$$\mathcal{M}_m = \pm \frac{5}{2}\sqrt{\left(\sum_q m_q \sin\alpha_q\right)^2 + \left(\sum_q m_q \cos\alpha_q\right)^2}$$

(4)

along the direction given by the unit vector $\boldsymbol{u}_m = \left(-\sum_q m_q \sin\alpha_q, \sum_q m_q \cos\alpha_q, 0\right)$ (*i.e.* lying in the triangles' planes).

Intra-ring (Equation (5)) and inter-ring (Equation (6)) dipolar coupling energies for the state $|m, M_{Cr}\rangle$ are ($\mu_B$ is the Bohr magneton):

$$E_{\text{dip},m}^{\text{intra}} = -\frac{\mu_0 \mu_B^2}{4\pi}\frac{125\,\eta^2}{3\sqrt{3}r^3}\sum_{q=1}^{3}(m_q m_{q+1} + m_{q'} m_{q'+1}),$$

(5)

$$E_{\text{dip},m}^{\text{inter}} = \frac{\mu_0 \mu_B^2}{4\pi}\frac{25\,\eta^2}{(r^2+h^2)^{\frac{3}{2}}}\left[2 - \frac{9r^2}{r^2+h^2}\right]\sum_{q=1}^{3} m_q(m_{q+1} + m_{q'-1})$$

$$-\frac{\mu_0 \mu_B^2}{4\pi}\frac{100\,\eta^2}{(4r^2+h^2)^{3/2}}\sum_{q=1}^{3} m_q m_{q'}.$$

(6)

Finally, $Dy^{III}$-$Cr^{III}$ dipolar coupling energy for $|m, M_{Cr}\rangle$ reads:

$$E_{\text{dip},m}^{\text{Dy-Cr}} = -\frac{\mu_0 \mu_B^2}{4\pi}\frac{40\,\eta}{(r^2+h^2/4)^{3/2}} M_{Cr}\mathcal{M}_m$$

(7)

The energy spectrum resulting from magnetic coupling can now be evaluated summing up Equations (3 -7), and is reported in Figure 7.

The dipolar coupling energies, Equations (5-7), contain no free-parameter, as they only depend on the specific set of quantum numbers $|m, M_{Cr}\rangle$, on the experimental geometrical parameters $r$ and $h$ (Fig. 6), and on the average ground state *ab initio* magnetic moment given by $10\eta\,\mu_B$, with $\eta$ =0.975. Interestingly, from Equation (5) it follows that intra-ring dipolar interactions always favor a toroidal texture on each triangle, penalizing the formation of a magnetic moment by an energy gap of ~4 cm$^{-1}$.

While inter-ring dipolar coupling is smaller than intra-ring coupling, due to larger separation between Dy$^{III}$ ions, from Equation (6) we learn that in fact this interaction splits FT and AFT states. In particular, the first term in Equation (6) stabilizes a FT (AFT) ground state if $h < r\sqrt{7/2}$ ($h > r\sqrt{7/2}$), while the second term (describing interactions between inversion-related centres) always favors FT coupling, and is stronger than the first term. In {Cr$^{III}$Dy$^{III}$$_6$}, $h > r\sqrt{7/2}$, hence the two terms are in competition, the first (second) favoring AFT (FT) coupling. Since the second term is larger, we find here that dipolar interactions stabilize a ferrotoroidic ground state in {Cr$^{III}$Dy$^{III}$$_6$} (see Figure 5).

Moreover, from Equation (6), we note that a structural design aimed at reducing the distance $h$ between the two triangles, will enhance ferrotoroidic coupling. We estimate the dipolar-induced FT/AFT splitting to be ~0.28 cm$^{-1}$. Note that, in our symmetric model, Dy-Cr dipolar interactions within FT and AFT states (Equation (7)) are exactly zero (since $\mathcal{M}_m = 0$ in FT and AFT states), leading to 8-fold degenerate FT and AFT manifolds. Thus in FT and AFT states the spin of Cr(III) is freely fluctuating, and FT/AFT toroidal excitations, involving the collective flipping of three Dy$^{III}$ spins, are fully determined by dipolar coupling. The energy gap of 3.3 cm$^{-1}$ reported in Figure 7 corresponds instead to the lowest magnetic excitation, obtained upon flipping a single Dy$^{III}$ spin. In the excited states the Cr(III) spin is blocked in the direction of the in-plane Dy$_6$ magnetic moment.

Considering now superexchange interactions Equation (3), we note that the dipolar-induced ferrotoroidic ground state will survive provided that the intra-ring Dy-Dy coupling is antiferromagnetic (i.e. $J_2 < 0$), or ferromagnetic but smaller than dipolar coupling. To estimate the two exchange coupling constants $J_1$ and $J_2$ appearing in Equation (3), we have employed DFT calculations, replacing the Dy$^{III}$ with Gd$^{III}$ ions in the X-ray structure. The computed coupling constants for the Cr$^{III}$-Gd$^{III}$ pairs were then rescaled by the ratio between the spin of Dy$^{III}$ ($S$=5/2) and that of Gd$^{III}$ ($S$=7/2), while Gd$^{III}$–Gd$^{III}$ were rescaled by the ratio between the square of $S$=5/2 and the square of $S$=7/2[33,34] We obtained $J_1 = -$ 0.08 cm$^{-1}$ (Cr$^{III}$-Dy$^{III}$ coupling) and $J_2 = -$ 0.043 cm$^{-1}$ (intra-ring Dy$^{III}$-Dy$^{III}$ coupling), indicating antiferromagnetic coupling. The estimated antiferromagnetic interaction within the {$Dy^{III}_3$} triangles reinforces the effect of the intra-ring dipolar coupling, leading to a toroidal moment on each isolated triangle.

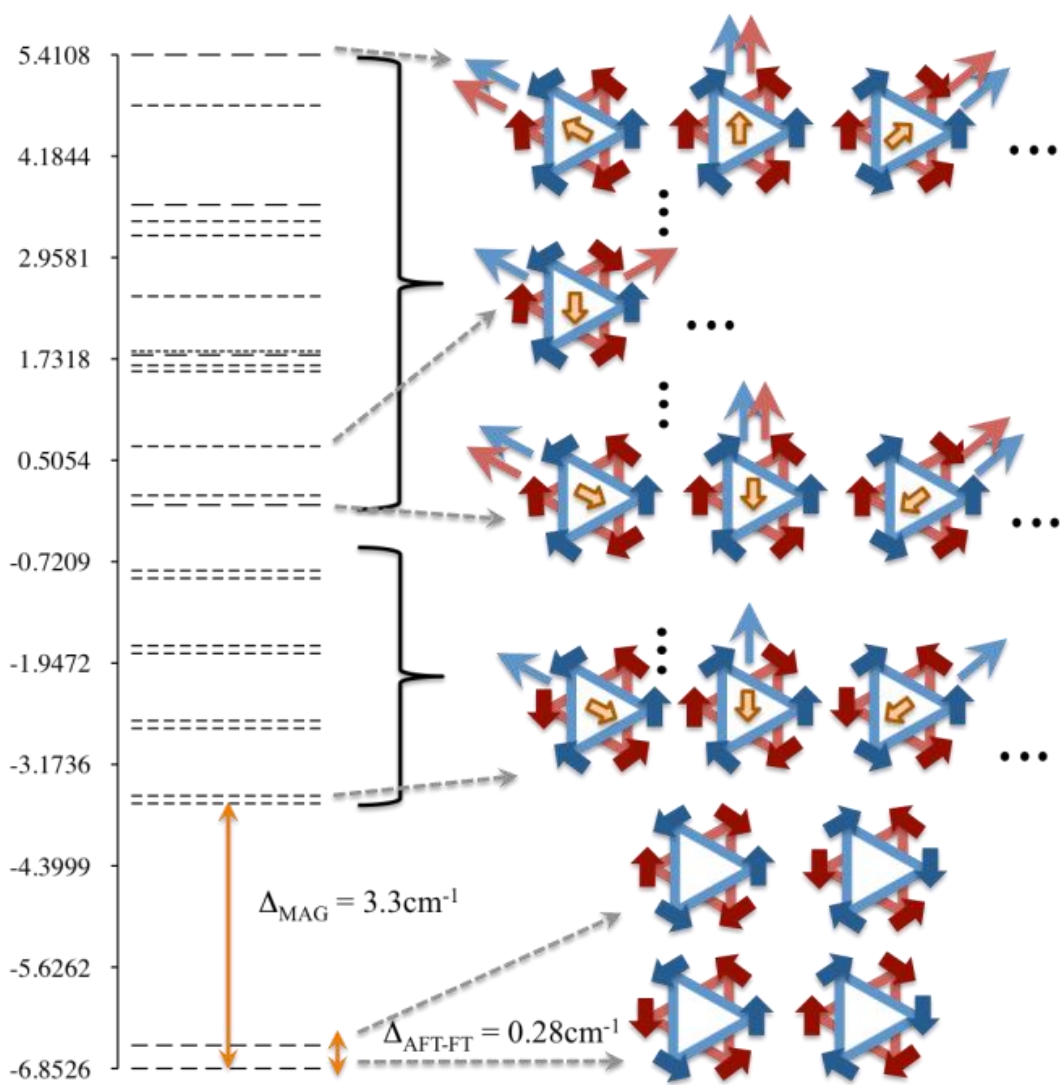

**Figure 7 | Energy spectrum describes magnetic relaxation and toroidal states.** Energy spectrum of {$Dy_6Cr$} calculated using the energy expressions Equations (3-7). Energies are in cm$^{-1}$. The number of dashes for each degenerate energy level indicates the number of degenerate states associated to that level. Besides a few energy levels it provides a graphical representation of one of the corresponding non-collinear Ising quantum states $|m, M_{Cr}\rangle$, where the red (blue) *thick* arrows at the Dy$^{III}$ sites indicate a +15/2 or -15/2 quantum number along the tangent direction in the lower (upper) triangular ring, a yellow thick arrow at the centre indicates the spin-projection of the Cr$^{III}$ ion relative to that of the Dy$^{III}$ ions, and a

resulting magnetic moment in the upper (lower) ring is indicated with *thin* blue (red) arrow. Gap between the ground FT state $|\pm\tau_1, \pm\tau_2, M_{Cr}\rangle$, and the first (AFT) $|\pm\tau_1, \mp\tau_2, M_{Cr}\rangle$ excited state are indicated.

While a degenerate FT quantum ground state is compatible with the inversion symmetry of the molecule, such symmetry is not compatible with a ferrotoroidically ordered phase. Thus upon FT ordering, a concomitant structural phase transition should occur to rid the crystal of the inversion centre, which, in turn, would allow the appearance of linear magneto-electric coupling. We now use the developed theoretical model to simulate the experimental molar susceptibility and magnetization. We report the results in Figures 2 and 8. While the magnetization at 2 K and for fields up to 5 T is expected to be dominated by the low-energy states described by our model Equations (1-2), the molar susceptibility will be dominated by these states only only at low temperatures, while at temperatures much higher than the coupling strength it will be dominated by the single-ion response, including the population of excited KDs. To reproduce the correct temperature dependence of $\chi_M T$ within our non-empirical model we therefore evaluated the molar susceptibility as $\chi T = (\chi T)_{LT} + (\chi T)_{HT} - (\chi_0 T)$, where $(\chi T)_{LT}$ is computed from our model Equations (1) and (2), $(\chi T)_{HT}$ is computed as a sum of the *ab initio* susceptibilities of the six $Dy^{III}$ ions and the central $Cr^{III}$ ion, while $(\chi_0 T) = 0.125 [6M(Dy)^2 + 4\times 3/2 \,(3/2+1)]$ is the uncoupled contribution to the Curie molar susceptibility (cm$^3$ K mol$^{-1}$) arising from the six $Dy^{III}$ ground KD's using the *ab initio* magnetic moment $M(Dy) = 9.75\mu_B$ and the ground quartet on $Cr^{III}$. The results of the calculations of $\chi T$ compared with the experimental data are reported in Figure 2a (orange line), which shows an excellent agreement between theory and experiment.

Note also that despite the absence of fitting parameters in our model, the theoretically calculated magnetization curve reported in Fig. 2b and Supplementary Fig. 7 (orange curve) reproduces very well the experimentally measured *M vs H* curve at 2 K (blue data points). In particular, the predicted FT and AFT low-energy states display a strongly reduced magnetic response, as is evident from the S-shape behavior of the *M vs H* curve. A uniform magnetic field does not interact with a toroidal moment, so that only the $Cr^{III}$ ion responds to the field for low *H*. As the Zeeman energy increases, at *H*~0.4 - 0.5T, a level crossing occurs (*vide infra*), corresponding to a magnetic excited state becoming the ground state, explaining the S-shape of the *M vs H* curve.

**Theoretical analysis of the zero-field hysteretic spin dynamics**

As a first attempt to interpret the single-crystal magnetization reported in Figure 8(a) (blue curve), measured at $T = 0.03$ K and a field sweep rate of 0.1 T/s, we calculated the equilibrium magnetization at the same temperature and magnetic field orientation (along the easy-axis, $y$-direction in Figure 6), also reported in Figure 8(a) (orange curve). Experimental and theoretical curves share a few common features, both displaying a rise of the magnetization for intermediate fields, separating two magnetization plateaus corresponding to low and high field regions. Theoretical and experimental plateaus are seen to coincide, with the theoretical low-field plateau corresponding to saturation of the free fluctuating Cr spin ($M_{SAT} \sim 3\mu_B$) within the FT ground state. According to the Zeeman spectrum in Figure 8(b) and 8(c), the steep magnetization step observed in the theoretical magnetization at the level-crossing field $B_{LC} \sim 0.4$ T can be identified with the switching of the ground state from the low-field weakly-magnetic FT Zeeman state $|\pm\tau, \pm\tau\rangle \equiv |\pm 1 \pm 1 \pm 1 \pm 1 \pm 1 \pm 1, -3/2\rangle$ (see Supplementary Figure 8(a)), to the high-field *onion* magnetic state $|+m, +m\rangle \equiv |+1 + 1 - 1 - 1 + 1 + 1, +3/2\rangle$, in which half of the Dy spins circulate clockwise, the other half anticlockwise (see Supplementary Figure 8(d)), adding up to a magnetic moment $M \sim 40\mu_B$ polarised along the field.

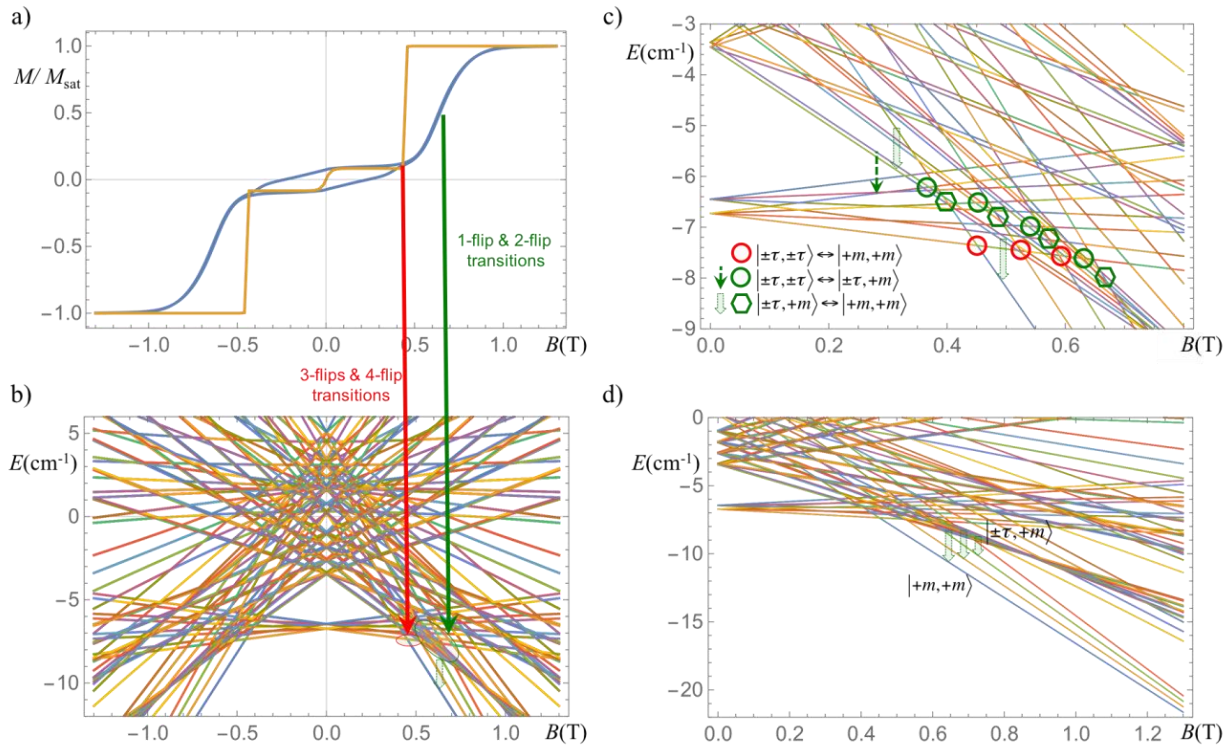

**Figure 8 | Interpretation of Single-crystal magnetization experiments.** a) Single-crystal magnetization experiment (blue curve) measured at $T = 0.03$ K and a sweep rate of 0.1 T/s for a magnetic field oriented along the in-plane easy-axis ($y$-axis in Figure 6), superimposed on the theoretical equilibrium magnetization (orange curve) computed for the same temperature and field direction; b) Energy levels as function of magnetic field (Zeeman spectrum) computed using the model presented in the text. The red (green) arrows connecting panels a) and b) associate the steep (smooth) raise of the theoretical (experimental) magnetization with slow/quasi-forbidden (fast/allowed) transition mechanisms occurring at level crossing between the two ground states (involving excited states) encircled in red (green) c) Magnification of the low-field region of the Zeeman spectrum, highlighting degeneracy points (level crossings) of states between which faster 1-flip and 2-flip tunneling transitions are allowed (green circles and hexagons), and also highlighting 3-flip or higher-flip processes that are essentially forbidden (red circles). Some of the allowed phonon emission processes are also indicated with dashed green arrows; d) High-field region of the Zeeman spectrum, where the four onion states antiferromagnetically coupled to Cr spin states become isolated from excited states, leading the system's magnetization to saturate.

As expected, the calculated equilibrium magnetization cannot reproduce inherently dynamical features observed in the experiments, as in the calculations transitions between energy levels are instantaneous on the timescale of the experiment. There are, in particular, two important differences between theory and experiment displayed in Figure 8(a): In the experiment a hysteresis loop opens up around the zero-field region between $-B_{LC}$ and $B_{LC}$ and the experimental magnetization step separating low and high-field plateaus is not abrupt, but occurs within a ~0.5 T field range starting at $B_{LC}$.

Thus, to further analyse the measured magnetization dynamics, we explicitly consider finite-rate transitions between Zeeman states. We are not interested in the microcscopic details of such transitions, other than these are driven by terms in the Hamiltonian that have so far been neglected (either because small, or because describing coupling with the surroundings), and other than broadly separating such processes in tunneling transitions characterised by coupling constants $\gamma_i$, active between degenerate energy levels, and phonon-induced transitions characterised by coupling constants $\Gamma_i$, from higher to lower energy states, with a probability that grows as the cube of the energy gap. The coupling constants $\gamma_i$ and $\Gamma_i$ are related to the matrix elements of the relevant Hamiltonian between initial and final states.

Crucially, we propose that a hierarchy should exist in the magnitude of $\gamma_i$ and $\Gamma_i$, based on the number of $Dy^{3+}$ spins that need to be flipped to change initial to final Zeeman states. We only consider in our model the 56 (out of 256) most relevant low-energy states involved in the relaxation dynamics, which are represented in Supplementary Figure 8. Particularly relevant are the low-field ground FT states $|\pm\tau, \pm\tau\rangle$, the high-field ground onion state $|+m, +m\rangle$, and the excited intermediate magnetic states $|\pm\tau, +m\rangle$ and $|+m, \pm\tau\rangle$ (Supplementary Figure 8(c)). Let us label the coupling constants $\gamma_i$ ($\Gamma_i$) as: $\gamma_{Cr}$ ($\Gamma_{Cr}$) for transitions that only flip the $Cr^{3+}$ spin; $\gamma_1$ ($\Gamma_1$) for transitions flipping only one $Dy^{3+}$ spin; $\gamma_2$ ($\Gamma_2$) for transitions involving two simultaneous spin-flipping events; $\gamma_3$ ($\Gamma_3$) for transitions involving three or more spin-flipping events. Thus the hierarchy we invoke here reads: $\gamma_{Cr} >> \gamma_1 > \gamma_2 > \gamma_3 \sim 0$, and $\Gamma_{Cr} >> \Gamma_1 > \Gamma_2 > \Gamma_3 \sim 0$.

Under such assumptions, we argue that both the zero-field hysteresis, and the smooth magnetization step observed in the experiment, can be explained by the fact that the direct $|\pm\tau, \pm\tau\rangle \leftrightarrow |+m, +m\rangle$ transition between the Zeeman ground states at level crossing $B_{LC} \sim 0.4T$ is essentially forbidden, as it involves at least the simultaneous flipping of three $Dy^{3+}$ spins ($\gamma_3, \Gamma_3 \sim 0$). Hence, on the timescale of field-sweeping, exchange of population between $|\pm\tau, \pm\tau\rangle$ and $|+m, +m\rangle$ can only be indirect, occurring via multi-step processes involving the excited intermediate states $|+m, \pm\tau\rangle$ and $|\pm\tau, +m\rangle$, also via the excited AFT states $|\pm\tau, \mp\tau\rangle$ (see Supplementary Figure 8(b)). This microscopic scenario is visualised in Figure 8(c), where a multitude of level crossings of states between which faster 1-flip and 2-flip transitions can occur are highlighted with green circles (tunneling) or green dashed arrows (phonon-mediated) ( $|\pm\tau, \pm\tau\rangle \leftrightarrow |\pm\tau, +m\rangle$ or $|\pm\tau, \pm\tau\rangle \leftrightarrow |+m, \pm\tau\rangle$ ), or highlighted with green hexagons (tunneling) and broad green arrows with a dashed contour (phonon-mediated) ($|\pm\tau, +m\rangle \leftrightarrow |+m, +m\rangle$ or $|+m, \pm\tau\rangle \leftrightarrow |+m, +m\rangle$). As illustrated in Figure 8(b) and 8(c), the existence of a broad range of magnetic fields around $B_{LC} \sim 0.4T$ for which fast transitions $|\pm\tau, \pm\tau\rangle \leftrightarrow |\pm\tau, +m\rangle$ can occur, followed, in higher fields, by fast phonon-relaxation from higher-energy intermediate to lower-energy onion states (see Figure 8(d)), provides a rationalization for the gradual rise of the experimental dynamical magnetization.

Moreover, to quantitatively investigate the origin of the observed zero-field hysteresis loop, we implement these ideas in a dynamical model based on generalised Pauli master equations describing the dissipative dynamics of the non-equilibrium thermal populations of the $CrDy_6$ states, coupled to an equilibrium reservoir of acoustic phonons at T = 0.03 K, and a source of random stray fields inducing incoherent tunneling between resonant energy levels. The states are obtained from our model Hamiltonians Equation (1) and Equation (2), also including a time-dependent Zeeman term describing the interaction with the sweeping magnetic field oriented along the easy-axis, at sweeping rate 0.1T/s, varying between $B = \pm 1.4$T (a triangle-wave signal is used). As outlined in the methods section, the relevant dissipative equations of motion for diagonal elements $\sigma_{ii}$ (populations) of the reduced density matrix $\sigma$, in the adiabatic approximation, are:

$$\dot{\sigma}_{ii}(t) = \sum_k \{W^{ph}_{k \to i}(t) + \Omega^{tun}_{k \leftrightarrow i}(t)\}\sigma_{kk}(t) - \sigma_{ii}(t)\sum_k \{W^{ph}_{i \to k}(t) + \Omega^{tun}_{k \leftrightarrow i}(t)\}$$

(8)

where $\Omega^{tun}_{k \leftrightarrow i}(t)$ and $W^{ph}_{i \to j}(t)$ are the time-dependent tunneling and phonon-induced transition rates given by Equations (12) and (10), respectively, proportional to the coupling constants $\gamma_i$ and $\Gamma_i$ fulfilling our proposed hierarchy $\gamma_{Cr} \gg \gamma_1 > \gamma_2 > \gamma_3 \sim 0$, and $\Gamma_{Cr} \gg \Gamma_1 > \Gamma_2 > \Gamma_3 \sim 0$. See methods section for a detailed discussion of the numerical choice of the parameters $\gamma_i$ and $\Gamma_i$.

We numerically solved Equation (8) for the time-dependent populations $\sigma(t)$ of the 56 $CrDy_6$ states considered here (see Supplementary Figure 8 and 10). The magnetization curve $M(t) = \text{Tr}[\sigma(t)M]$ is then computed ($M$ is the magnetic moment operator along the field), parametrically plotted *vs* the sweeping field, and reported in Figure 9 with the experimental magnetization. Quite remarkably, the opening of the hysteresis in the field region between $+B_{LC}$ and $-B_{LC}$ is captured by our model, together with the narrowing and closing of the hysteresis loop at fields $\sim B_{LC}$. Also, the simulated dynamical magnetization now displays a smooth increase between the low and high field "plateaus", despite the low temperature, on account of the cascade of indirect transitions between the low-field (ferrotoroidic) and high-field (onion) ground states mediated by intermediate excited magnetic states.

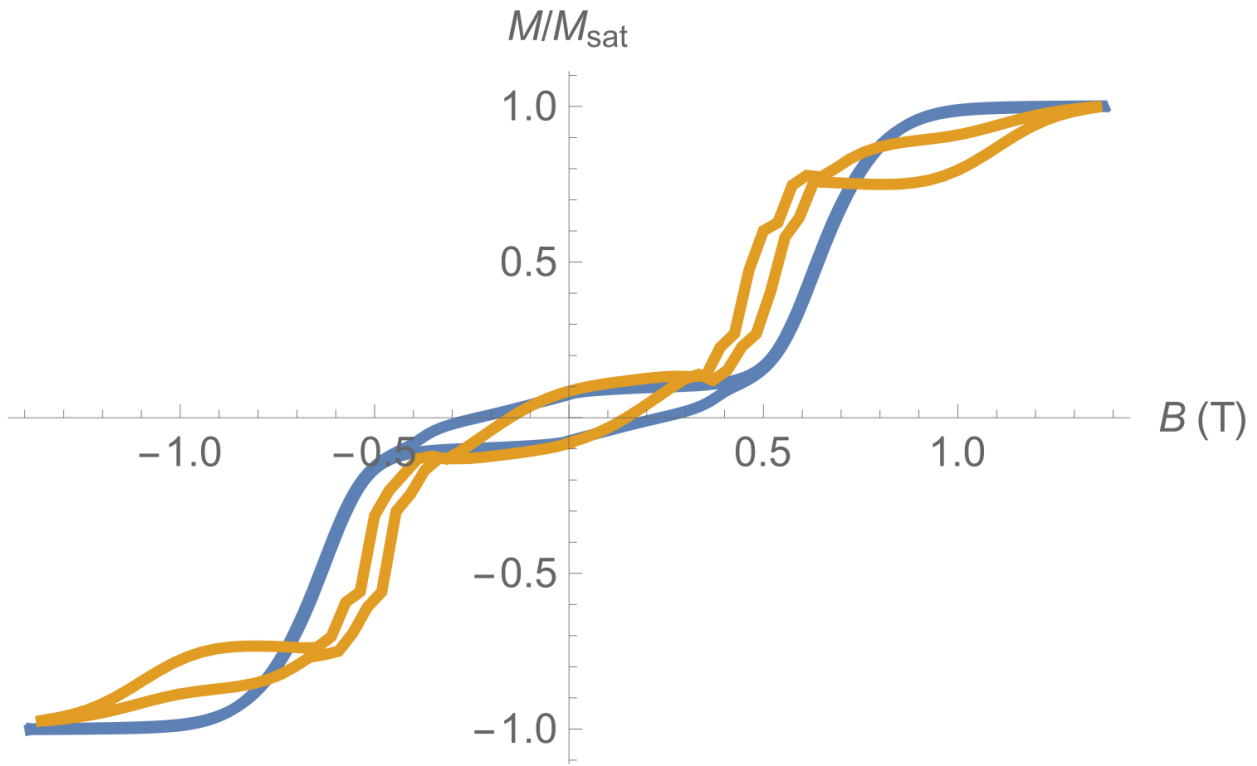

**Figure 9 | Simulation of single-crystal magnetization**. Single-crystal experimental magnetization (blue curve) measured at $T = 0.03$ K and a sweep rate of 0.1 T/s for a magnetic field oriented parallel to the triangles' planes and along the easy-axis ($y$ axis in Figure 6), superimposed on the simulated dynamical magnetization at the same temperature, sweep rate and field orientation, by solving Equation (8) on the basis of 56 out of the 256 states obtained from our model and reported in Supplementary Figure 8, for the following numerical values of the transition rates appearing in the equations: $\Gamma_{Cr} = 10^5$ Hz/(cm$^{-1}$)$^3$ >> $\Gamma_1 = 10^{-7} \times \Gamma_{Cr}$ > $\Gamma_2 = 10^{-3} \times \Gamma_1$ , and $\gamma_{Cr} = 10^{16}$ Hz$^2$ >> $\gamma_1 = 10^{12}$ Hz$^2$ > $\gamma_2 = 10^{-3} \times \gamma_1$.

There are, of course, shortcomings in our simulation. For instance, the zero-field hysteresis loop, after closing down as in the experiment, opens up again at larger fields, a feature that is only marginally present in the experiment. Also, the theoretical magnetization step covers a smaller field range than the experimental one. While we cannot exclude that a more thrrorough exploration of the parameter space might improve the fitting, we expect that these shortcomings will be partially overcome if all 256 states were included in the simulation, as the additional excited states would participate to the multi-step magnetization relaxation, further widening the field range covered by the magnetization step. Further discussion of our dynamical model, including a partitioning of the contributions to $M(t)$ from individual states, is reported in the Supplementary Note 5.

In summary, a heptanuclear {$Cr^{III}Dy^{III}_6$} complex has been synthesised and structurally characterized. Experimental evidence in conjunction with theoretical calculations reveal and explain the presence of both single-molecule magnet and single-molecule toroidic behaviour. The toroidal states in the individual {$Dy^{III}_3$} triangles are found to be ferrotoroidically coupled. We note here that the $Cr^{III}$ ion does not play any fundamental role in the predicted ferrotoroidic coupling and the coupling between the two toroidal wheels can in fact be fully explained solely in terms of dipolar interactions, which depend solely on the structural parameters of the complex. The fundamental structural elements influencing the strength of dipolar ferrotoroidic coupling are the staggered arrangement of the two triangles with respect to each other, and the distance $h$ between the two wheels, which we found not to be optimal for maximizing ferrotoroidic coupling (i.e. $h > r\sqrt{7/2}$). The design and synthesis of similar {$M^{III}Dy^{III}_6$} complexes, maintaining the staggered arrangement of the two triangular units, but featuring a smaller, even diamagnetic ion, is expected to lead to a smaller inter-ring distance $h$, or even to $h < r\sqrt{7/2}$, thus according to Equation (6), to a stronger ferrotoroidic coupling. This route is currently being explored in our labs. Importantly, our findings indicate, for the first time, how coupling between toroidal moments can be manipulated by structural design. Finally, for the first time, the experimental single-crystal magnetization dynamics of a polynuclear Dy complex, displaying zero-field opening of a hysteresis loop, is simulated via a theoretical dynamical model, showing that the ferrotoroidic ground state plays a pivotal role in hindering the flipping of magnetic onion states in a sweeping field, thus slowing down the zero-field magnetic relaxation.

## Methods

**Synthesis of [$Dy^{III}_6Cr^{III}(OH)_8(ortho$-tol$)_{12}(MeOH)_5(NO_3)$]·3MeOH (1).** $Cr(NO_3)_3$·$9H_2O$ (0.4 g, 1 mmol) and $Dy(NO_3)_3$·$6H_2O$ (0.22 g, 0.5 mmol) were dissolved in MeCN (20 mL), followed by the addition *ortho*-toluic acid (0.14 g, 1.0 mmol) and triethylamine (0.55 mL, 4.0 mmol), which resulted in a pale purple solution. This solution was stirred for 4 hours after which time the solvent was removed to give a purple oil. The oil was re-dissolved in MeOH/$^i$PrOH (1:1) and the solution allowed to slowly evaporate. Within 10-15 days' pale purple crystals of 1 had appeared, in approximate yield of 25% (crystalline product). Microanalysis for $CrDy_6C_{104}H_{124}NO_{43}$: Expected (found); C 40.25 (39.86), H 4.02 (3.86), N 0.45 (0.62).

The synthesis reaction was carried out under aerobic conditions. Chemicals and solvents were obtained from commercial sources and used without further purification.

**X-ray crystallography**

X-ray measurements for **1** were performed at 100(2) K at the Australian synchrotron MX1 beam-line.[45] The data collection and integration were performed within Blu-Ice[46] and XDS[47] software programs. Compound **1** was solved by direct methods (SHELXS-97),[48] and refined (SHELXL-97)[49] by full least matrix least-squares on all $F^2$ data within X-Seed[50] and OLEX-2 GUIs.[51] Crystallographic data and refinement parameters are summarized in Supplementary Table 1.

**Magnetic measurements**

The magnetic susceptibility measurements were carried out on a Quantum Design SQUID magnetometer MPMS-XL 7 operating between 1.8 and 300 K for dc-applied fields ranging from 0 – 5 T. Microcrystalline samples were dispersed in Vaseline in order to avoid torquing of the crystallites. The sample mulls were contained in a calibrated gelatine capsule held at the centre of a drinking straw that was fixed at the end of the sample rod. Alternating current (ac) susceptibilities were carried out under an oscillating ac field of 3.5 Oe and frequencies ranging from 0.1 to 1500 Hz.

**EPR Instrumentation**

The X-band measurements were made on Bruker spectrometers, at IIT Bombay, with a helium gas-flow cryostat. The X-band measurements were carried out at 5, 10 and 20 K.

**Computational Details**

Even though there are only three crystallographic nonequivalent $Dy^{III}$ centers, we performed the calculations on all six $Dy^{III}$ ions to determine the direction of the local anisotropy axis. Using MOLCAS 7.8[43], *ab initio* calculations were performed on the six $Dy^{III}$ ions using the crystal structure of **1**. The structure of the modelled Dy fragment employed for calculation is shown in Supplementary Fig. 5, where the neighbouring $Dy^{III}$ ions are replaced with diamagnetic $Lu^{III}$ ions and the $Cr^{III}$ ion replaced with $Sc^{III}$ ion. We have employed this methodology to study a number of $Dy^{III}/Er^{III}$ SMMs.[52-55] The relativistic effects are taken into account based on the Douglas−Kroll Hamiltonian.[56] The spin-free eigen-states are achieved by the Complete Active Space Self-Consistent Field (CASSCF) method.[57] We have employed the [ANO-RCC...8s7p5d3f2g1h.] basis set[58] for the Dy atoms, the [ANO-RCC...3s2p.] basis set for the C atoms, the [ANO-RCC...2s.] basis set for H atoms, the [ANO-RCC...3s2p1d.] basis set for the N atoms, the [ANO-RCC...4s3p1d.] basis set for the Sc atom, the [ANO-RCC...5s4p2d1f.] basis set for the La atom and the [ANO-RCC...3s2p1d.] basis set for the O atoms. First we performed the CASSCF calculation including nine electrons across seven 4f orbitals of the $Dy^{3+}$ ion. With this active space, we have computed 21 roots in the Configuration Interaction (CI) procedure. After computing these excited states, we have mixed all roots using RASSI-SO[59] and spin-orbit coupling is considered within the space of calculated spin-free eigenstates. Moreover, we have considered these computed SO states into the SINGLE_ANISO[60] program to compute the *g*-tensors. The Dy$^{III}$ ion has eight low-lying Kramer's doublets for which the anisotropic *g*-tensors have been computed. The Cholesky decomposition for two electron integrals is employed throughout our calculations. We have extracted the crystal field parameters using the SINGLE_ANISO code as implemented in MOLCAS 7.8.

DFT calculations were performed using the B3LYP functional[61] with the *Gaussian 09* suite of programs.[62] To estimate the exchange constant between Cr$^{III}$-Dy$^{III}$ and Dy$^{III}$-Dy$^{III}$ ions, the dysprosium ions were replaced with the spin-only Gd$^{III}$ ions in order to investigate the exchange interaction between the Dy$^{III}$ ions, which was then rescaled to the spin of dysprosium ions. We have used the LanL2DZ ECP basis set for Cr,[63-64] the double-zeta quality basis set employing Cundari-Stevens (CS) relativistic effective core potential on Gd atom,[65] 6-31G* basis set for the rest of the atoms. The DFT calculations combined with the Broken Symmetry (BS) approach[66] have been employed to compute the magnetic exchange *J* value. The actual energy spectrum and wavefunction and magnetic texture calculations, together with simulation of the magnetic measurements, was carried out by inserting the *ab initio* data into a model intramolecular magnetic coupling Hamiltonian, the development of which is described in the theoretical analysis section (see above).

**Method of simulation of the magnetization dynamics**

As is well known[67], in the derivation of the Pauli equations for the diagonal reduced density matrix $\sigma$ from the Liouville-Von Neumann equations for the total density matrix describing the coupled system-phonon reservoir it is expedient to switch to the interaction picture, so that the unperturbed quantum system and reservoir Hamiltonians do not explicitly appear in the transformed equations of motion. The equations of motion in the interaction picture are thus propagated only by the spin-phonon coupling Hamiltonian, which in second order and in the Born-Markov limit generates the dissipative relaxation dynamics of the reduced density matrix. In our particular case, where time-dependence arises both from the random dissipative relaxation fields, and from the periodic time-dependence of the Zeeman Hamiltonian, a useful interaction picture can still be achieved by including the time-dependent Zeeman Hamiltonian in the zero-th order Hamiltonian $H_0(t) \equiv H_0$ together with Hamiltonian Equations (1) and (2). To avoid the complications related to the resulting time-dependence of $H_0$ which would imply a zeroth order time-evolution operator only defined exactly via time-ordered products, we assume valid the adiabatic approximation, so that at any time the system is assumed to be described by the eigenfunctions $|i(t)\rangle$ and eigenvalues $E_i(t)$ of the time-dependent Hamiltonian $H_0$, so that $H_0(t)|i(t)\rangle = E_i(t)|i(t)\rangle$, and the time evolution operator can be simplified as an exponential operator that remains diagonal in the basis of the 256 system's eigenfunctions $|i(t)\rangle$ at any time. Using well-known approximations to avoid at each given time the integration of eigenvalues at all previous times, as described for instance in the paper[68] and within the secular approximation[67], two decoupled sets of equations of motion are obtained for the elements of the reduced density matrix $\sigma_{ij} = \langle i|\sigma|j\rangle$ between the adiabatic eigenstates of $H_0$, one for the diagonal (populations, $i = j$) and one for the off-diagonal (coherences, $i \neq j$) elements of the CrDy$_6$ reduced density matrix $\sigma$, which can be written in the usual Schrödinger's picture as:

$$\dot{\sigma}_{ij}(t) = -i/\hbar \langle i|[H_0(t), \sigma(t)]|j\rangle + \delta_{ij} \sum_k W_{k\to i}(t)\sigma_{kk}(t) - \mu_{ij}(t)\sigma_{ij}(t)$$

(9)

where $\dot{\sigma}_{ij} \equiv d\sigma_{ij}/dt$, $[H_0, \sigma]$ is the commutator between the operators $H_0$ and $\sigma$, $\delta_{ij}$ is a Kronecker delta, $W_{l\to m}(t)$ is the transition rate from eigenstate $|l(t)\rangle$ to eigenstate $|m(t)\rangle$, which become themselves time-dependent because of the time-dependent Zeeman field, both via the Zeeman eigenvalues $E_l(t)$, but in principle also via the eigenstates $|l(t)\rangle$, as Cr spin states are now coupled by the magnetic field. Finally, we have $\mu_{ij}(t) = 1/2 \sum_k (W_{i\to k}(t) + W_{j\to k}(t))$.

The details of the spin-phonon coupling Hamiltonian contributing to the transition rates for this polynuclear system coupled to its crystal phonons represents in principle a formidable problem, whose detailed analysis goes beyond the scope of the current paper. Hence we describe here the coupling of the CrDy$_6$ states to an idealized equilibrium acoustic phonon reservoir in terms of well-known phenomenological transition rates $W_{l\to m}(t)$ obtained within the Debye model[2] and given by:

$$W_{i\to j}^{ph}(t) = \Gamma_{ij} \frac{\left(E_i(t) - E_j(t)\right)^3}{\exp\left[\left(E_i(t) - E_j(t)\right)/k_B T\right] - 1}, \qquad \Gamma_{ij} = \Gamma_{Cr}, \Gamma_1, \Gamma_2$$

(10)

where $E_i(t)$ are now the time-dependent Zeeman energies of the CrDy$_6$ quantum states, while $\Gamma_{ij}$ are numerical parameters measuring the transition rate $\Gamma_{Cr}$, or $\Gamma_1$, or $\Gamma_2$, according to how many spin-flips connect states $|i(t)\rangle$ and $|j(t)\rangle$ (we set $\Gamma_{ij} = 0$ if the two states are connected by 3 or more spin flips).

Finally, in order to account for the relaxation dynamics associated to quantum tunneling processes induced by e.g. random stray magnetic fields produced by the fluctuations of nuclear dipole moments or neighboring molecular magnetic moments, we follow the approach proposed by Leuenberger and Loss[69] [see in particular the derivation of their Eq. (35) and (36)], and further correct the zero-th order Hamiltonian $H_0$ in Equation (9) with an additional term $V$ describing tunneling between the Dy$^{3+}$ and Cr$^{3+}$ magnetic states via e.g. coupling to fluctuating magnetic fields arising from neighboring spins. We note in fact that in Equation (9), the diagonal matrix elements of the commutator between the eigenstates of $H_0$ are exactly zero in the adiabatic approximation, and thus the differential equations for the populations are there decoupled from those for the coherences. If on the other hand we now introduce the tunneling operator $V$, so that in Equation (8) $H_0(t) \to H_0(t) + V$, since $V$ by definition has non-zero off-diagonal matrix elements between the eigenstates of $H_0$, this correction will re-introduce coupling between populations and coherences in Equation (9). Assuming a steady-state approximation for the coherences, so that $\dot{\sigma}_{ij}(t) \approx 0$ on the timescale over which the populations $\sigma_{kk}(t)$ display appreciable changes, Equation (9) can be transformed in a system of differential equations for the populations only, a set of generalized Pauli equations that reads (this is Equation (8), reported here for convenience):

$$\dot{\sigma}_{ii}(t) = \sum_k \left\{W_{k\to i}^{ph}(t) + \Omega_{k\leftrightarrow i}^{tun}(t)\right\} \sigma_{kk}(t) - \sigma_{ii}(t) \sum_k \left\{W_{i\to k}^{ph}(t) + \Omega_{k\leftrightarrow i}^{tun}(t)\right\}$$

(11)

where the time-dependent incoherent tunneling transition rates $\Omega_{k \leftrightarrow i}^{tun}(t)$ are given by:

$$\Omega_{k \leftrightarrow i}^{tun}(t) = \gamma_{ki} \frac{\mu_{ij}(t)}{\omega_{ij}^2(t) + |\mu_{ij}(t)|^2} \approx \gamma_{ki} \frac{\lambda}{\omega_{ij}^2(t) + \lambda^2}, \quad \text{with} \gamma_{ki} = \gamma_{Cr}, \gamma_1, \gamma_2$$

(12)

where $\omega_{ij}(t) = \big(E_i(t) - E_j(t)\big)/\hbar$. We note that Equation (12) is in fact the equation of a Lorentzian lineshape with a maximum at the resonance $\omega_{ij}(t) = 0$, thus describing incoherent tunneling processes between Zeeman states in proximity of level crossing. While in principle this derivation predicts via Equation (12) that the broadening of the Lorentzian lineshape should also be treated as time-dependent, we choose here to fix the broadening as a constant parameter $\lambda$ to simplify our dynamical model.

The task of solving Equation (11) is computationally not trivial, and to make the calculations faster and more stable, instead of including all the 256 states of the $CrDy_6$ system arising from our low-energy model, we decided to include only the lowest energy states over the magnetic field range explored, which beside the sixteen ferrotoroidic and antiferrotoroidic states, they correspond to Dy-based magnetic states with anisotropy axes aligned along the sweeping field. These states correspond in fact the configurations shown in Supplementary Figure 8. We report in Supplementary Figure 10 the 56 Zeeman levels entering our dynamical model Equation (11), as function of the magnetic field, which can be compared with the full plot of the 256 Zeeman levels in Figure 8(b-d) in the main text. Note that in principle we have at least seven free parameters entering Equation (11): the three spin-phonon relaxation rates $\Gamma_{Cr}$, $\Gamma_1$, and $\Gamma_2$, with $\Gamma_{Cr} \gg \Gamma_1 > \Gamma_2$, the three squares of tunneling relaxation rates $\gamma_{Cr}$, $\gamma_1$, and $\gamma_2$ with $\gamma_{Cr} \gg \gamma_1 > \gamma_2$, and the Lorentzian broadening $\lambda$. The actual maximal relaxation rate at the Lorentzian maximum (i.e. at exact level crossing) is in fact given by $\gamma_k/\lambda$, with $\gamma_k$ corresponding to the *square* of the tunnelling splitting at level crossing expressed in Hz. Some of these parameters can in fact be fixed within reasonable ranges. For instance, from Ref.[70] we learn that the typical range of values for $\Gamma_{Cr}$, i.e. the spin-phonon relaxation rate for simple paramagnetic ions, is in the range $3 \times 10^3$ to $3 \times 10^5$ Hz/(cm$^{-1}$)$^3$. We further assume that fluctuating stray fields are of the order of 1mT, given that the magnitude of the non-fluctuating dipolar field induced at any $Cr^{3+}$ or $Dy^{3+}$ site within one molecule varies between ~8mT (inter-wheel interactions) and ~200mT (intra-wheel interactions), and that due to the strongly anisotropic and non-collinear character of the local magnetic moments, orientational effects will greatly reduce these fields, which can only induce tunneling when oriented perpendicular to the local anisotropy axes. Considering that the transition magnetic moment between $Cr^{3+}$ spin states whose $M_S$ quantum number differ by one unit is ~$2\mu_B$, we get the following rough estimation for $\gamma_{Cr} \sim (2\mu_B \times 1mT)^2/\hbar^2 \sim 10^{16} Hz^2$ (corresponding to a maximal tunneling frequency in the absence of broadening of ~ 0.1GHz). Also, from our single-ion ab initio calculations for the $Dy^{3+}$ ions in $CrDy_6$, we find that the average value of the transition matrix element between the two components $M_J \sim \pm 15/2$ of the ground Kramers doublet on each ion is of the order of ~$10^{-2}\mu_B$. Given that the 1-flip tunneling transitions only involve one $Dy^{3+}$ ion at the time, we obtain $\gamma_1 \sim (10^{-2}\mu_B \times 1mT)^2/\hbar^2 \sim 10^{11} \div$ $10^{12}$ Hz$^2$ (i.e. a maximal tunneling frequency of 1MHz). While these rough considerations reduce the possible values for three of the seven free parameters, we still need to arbitrarily fix four of them: $\Gamma_1$, $\Gamma_2$, $\gamma_2$ and $\lambda$.

Given the approximate nature of the model, it is not our aim to attempt a fully satisfying fitting of the remaining four parameters. Thus guided by the relations $\Gamma_{Cr} = 10^5$ Hz/(cm$^{-1}$)$^3 \gg \Gamma_1 > \Gamma_2$, and $\gamma_{Cr} = 10^{16}$ Hz$^2 \gg \gamma_1 = 10^{12}$ Hz$^2 > \gamma_2$, we found that a reasonably good agreement between the calculated magnetization $M(t) = \text{Tr}[\sigma(t)M]$ and experimental magnetization can be obtained for $\Gamma_1 = 10^{-7} \times \Gamma_{Cr}$, $\Gamma_2 = 10^{-3} \times \Gamma_1$, $\gamma_2 = 10^{-3} \times \gamma_1$, and $\lambda = 10^{10}$ Hz, which are the parameters used to obtain Supplementary Figure 11.

**Data availability**

The X-ray crystallographic coordinates for structure reported in this study have been deposited at the Cambridge Crystallographic Data Centre (CCDC), under deposition number 1435033. These data can be obtained free of charge from the Cambridge Crystallographic Data Centre via www.ccdc.cam.ac.uk/data_request/cif.

## References

1. Christou, G., Gatteschi, D., Hendrickson, D. N. & Sessoli, R. Single-molecule magnets. *MRS Bull* **25**, 66-71 (2000).
2. D. Gatteschi, R. Sessoli & Villain, J. *Molecular Nanomagnets*. (Oxford University Press, Oxford, 2006).
3. Gatteschi, D. Molecular Magnetism: A basis for new materials. *Adv. Mater.* **6**, 635-645 (1994).
4. Gatteschi, D. & Sessoli, R. Quantum Tunneling of Magnetization and Related Phenomena in Molecular Materials. *Angew. Chem. Int. Ed.* **42**, 268-297 (2003).
5. Sessoli, R., Gatteschi, D., Caneschi, A. & Novak, M. A. Magnetic Bistability in a Metal-Ion Cluster. *Nature* **365**, 141-143 (1993).
6. Sessoli, R. *et al.* High-spin molecules: [Mn$_{12}$O$_{12}$(O$_2$CR)$_{16}$(H$_2$O)$_4$]. *J. Am. Chem. Soc.* **115**, 1804-1816 (1993).
7. Gutlich, P. & Goodwin, H. A. *Spin Crossover in Transition Metal Compounds I.*, (Springer, 2004).
8. Tang, J. *et al.* Dysprosium Triangles Showing Single-Molecule Magnet Behavior of Thermally Excited Spin States. *Angew. Chem. Int. Ed.* **45**, 1729-1733 (2006).
9. Chibotaru, L. F., Ungur, L. & Soncini, A. The Origin of Nonmagnetic Kramers Doublets in the Ground State of Dysprosium Triangles: Evidence for a Toroidal Magnetic Moment. *Angew. Chem. Int. Ed.* **47**, 4126-4129 (2008).
10. Lin, S.-Y. *et al.* Coupling Dy$_3$ Triangles to Maximize the Toroidal Moment. *Angew. Chem. Int. Ed.* **51**, 12767-12771 (2012).
11. Novitchi, G. *et al.* Heterometallic Cu$^{II}$/Dy$^{III}$ 1D chiral polymers: chirogenesis and exchange coupling of toroidal moments in trinuclear Dy$_3$ single molecule magnets. *Chem. Sci.* **3**, 1169-1176 (2012).
12. Soncini, A. & Chibotaru, L. F. Toroidal magnetic states in molecular wheels: Interplay between isotropic exchange interactions and local magnetic anisotropy. *Phys. Rev. B* **77**, 220406 (2008).
13. Ungur, L., Lin, S.-Y., Tang, J. & Chibotaru, L. F. Single-molecule toroics in Ising-type lanthanide molecular clusters. *Chem. Soc. Rev.* **43**, 6894-6905 (2014).
14. Ungur, L., Van den Heuvel, W. & Chibotaru, L. F. Ab initio investigation of the non-collinear magnetic structure and the lowest magnetic excitations in dysprosium triangles. *New J. Chem.* **33**, 1224-1230 (2009).
15. Gysler, M. *et al.* Multitechnique investigation of Dy$_3$ - implications for coupled lanthanide clusters. *Chem. Sci.* **7**, 4347-4354 (2016).
16. Gatteschi, D., Sessoli, R. & Sorace, L. in *Handbook on the Physics and Chemistry of Rare Earths* Vol. Volume 50 (eds G. Bünzli Jean-Claude & K. Pecharsky Vitalij) 91-139 (Elsevier, 2016).
17. Luzon, J. *et al.* Spin Chirality in a Molecular Dysprosium Triangle: The Archetype of the Noncollinear Ising Model. *Phys. Rev. Lett.* **100**, 247205 (2008).
18. Hewitt, I. J. *et al.* Coupling Dy$_3$ Triangles Enhances Their Slow Magnetic Relaxation. *Angew. Chem. Int. Ed.* **49**, 6352-6356 (2010).

19. Hussain, B., Savard, D., Burchell, T. J., Wernsdorfer, W. & Murugesu, M. Linking high anisotropy $Dy_3$ triangles to create a $Dy_6$ single-molecule magnet. *Chem. Commun.*, 1100-1102 (2009).
20. Katsura, H., Nagaosa, N. & Balatsky, A. V. Spin Current and Magnetoelectric Effect in Noncollinear Magnets. *Phys. Rev. Lett.* **95**, 057205 (2005).
21. Soncini, A. & Chibotaru, L. F. Molecular spintronics using noncollinear magnetic molecules. *Phys. Rev. B* **81**, 132403 (2010).
22. Trif, M., Troiani, F., Stepanenko, D. & Loss, D. Spin-Electric Coupling in Molecular Magnets. *Phys. Rev. Lett.* **101**, 217201 (2008).
23. Das, C. *et al.* Single-Molecule Magnetism, Enhanced Magnetocaloric Effect, and Toroidal Magnetic Moments in a Family of $Ln_4$ Squares. *Chem. Eur. J.* **21**, 15639-15650 (2015).
24. Guo, P.-H. *et al.* The First {$Dy_4$} Single-Molecule Magnet with a Toroidal Magnetic Moment in the Ground State. *Inorg. Chem.* **51**, 1233-1235 (2012).
25. Popov, A. I., Plokhov, D. I. & Zvezdin, A. K. Magnetoelectricity of single molecular toroics: The {$Dy_4$} ring cluster. *Phys. Rev. B* **94**, 184408 (2016).
26. Ungur, L. *et al.* Net Toroidal Magnetic Moment in the Ground State of a {$Dy_6$}-Triethanolamine Ring. *J. Am. Chem. Soc.* **134**, 18554-18557 (2012).
27. Kaelberer, T., Fedotov, V. A., Papasimakis, N., Tsai, D. P. & Zheludev, N. I. Toroidal Dipolar Response in a Metamaterial. *Science* **330**, 1510-1512 (2010).
28. Chilton, N. F., Langley, S. K., Moubaraki, B. & Murray, K. S. Synthesis, structural and magnetic studies of an isostructural family of mixed 3d/4f tetranuclear 'star' clusters. *Chem. Commun.* **46**, 7787-7789 (2010).
29. Langley, S. K. *et al.* Heterometallic Tetranuclear [$Ln^{III}_2Co^{III}_2$] Complexes Including Suppression of Quantum Tunneling of Magnetization in the [$Dy^{III}_2Co^{III}_2$] Single Molecule Magnet. *Inorg. Chem.* **51**, 11873-11881 (2012).
30. Li, M. *et al.* A Family of 3d-4f Octa-Nuclear [$Mn^{III}_4Ln^{III}_4$] Wheels (Ln = Sm, Gd, Tb, Dy, Ho, Er, and Y): Synthesis, Structure, and Magnetism. *Inorg. Chem.* **49**, 11587-11594 (2010).
31. Schray, D. *et al.* Combined Magnetic Susceptibility Measurements and $^{57}Fe$ Mössbauer Spectroscopy on a Ferromagnetic {$Fe^{III}_4Dy_4$} Ring. *Angew. Chem. Int. Ed.* **49**, 5185-5188 (2010).
32. Vignesh, K. R., Langley, S. K., Moubaraki, B., Murray, K. S. & Rajaraman, G. Large Hexadecametallic {$Mn^{III}$–$Ln^{III}$} Wheels: Synthesis, Structural, Magnetic, and Theoretical Characterization. *Chem. Eur. J.* **21**, 16364-16369 (2015).
33. Langley, S. K. *et al.* A {$Cr^{III}_2Dy^{III}_2$} Single-Molecule Magnet: Enhancing the Blocking Temperature through 3d Magnetic Exchange. *Angew. Chem. Int. Ed.* **52**, 12014-12019 (2013).
34. Langley, S. K. *et al.* Modulation of slow magnetic relaxation by tuning magnetic exchange in {$Cr_2Dy_2$} single molecule magnets. *Chem. Sci.* **5**, 3246-3256 (2014).
35. Langley, S. K., Wielechowski, D. P., Moubaraki, B. & Murray, K. S. Enhancing the magnetic blocking temperature and magnetic coercivity of {$Cr^{III}_2Ln^{III}_2$} single-molecule magnets via bridging ligand modification. *Chem. Commun.* **52**, 10976-10979 (2016).
36. Rinck, J. *et al.* An Octanuclear [$Cr^{III}_4Dy^{III}_4$] 3d–4f Single-Molecule Magnet. *Angew. Chem. Int. Ed.* **49**, 7583-7587 (2010).
37. Cirera, J., Ruiz, E. & Alvarez, S. Shape and Spin State in Four-Coordinate Transition-Metal Complexes: The Case of the $d^6$ Configuration. *Chem. Eur. J.* **12**, 3162-3167 (2006).
38. Pinsky, M. & Avnir, D. Continuous Symmetry Measures. 5. The Classical Polyhedra. *Inorg. Chem.* **37**, 5575-5582 (1998).
39. Cadiou, C. *et al.* Studies of a nickel-based single molecule magnet: resonant quantum tunnelling in an S=12 molecule. *Chem. Commun.*, 2666-2667 (2001).
40. Wang, D. & Hanson, G. R. New methodologies for computer simulation of paramagnetic resonance spectra. *Appl. Magn. Reson.* **11**, 401-415 (1996).
41. Wang, D. & Hanson, G. R. Extreme Rhombic Distortion in Electron Paramagnetic Resonance and a Superposition Model Account. *J. Magn. Reson. Ser A* **118**, 1-6 (1996).
42. Moreno Pineda, E. *et al.* Direct measurement of dysprosium(III)···dysprosium(III) interactions in a single-molecule magnet. *Nat. Commun.* **5**, 5243 (2014).
43. Aquilante, F., Pedersen, T. B., Veryazov, V. & Lindh, R. MOLCAS—a software for multiconfigurational quantum chemistry calculations. *WIREs Comput. Mol. Sci.* **3**, 143-149 (2013).
44. Lines, M. E. Orbital Angular Momentum in the Theory of Paramagnetic Clusters. *J. Chem. Phys.* **55**, 2977-2984 (1971).

45. Cowieson, N. P. *et al.* MX1: a bending-magnet crystallography beamline serving both chemical and macromolecular crystallography communities at the Australian Synchrotron. *J. Synchrotron Radiat.* **22**, 187-190 (2015).
46. McPhillips, T. M. *et al.* Blu-Ice and the Distributed Control System: software for data acquisition and instrument control at macromolecular crystallography beamlines. *J. Synchrotron Radiat.* **9**, 401-406 (2002).
47. Kabsch, W. Automatic processing of rotation diffraction data from crystals of initially unknown symmetry and cell constants. *J. Appl. Crystallogr.* **26**, 795-800 (1993).
48. Sheldrick, G. A short history of SHELX. *Acta Crystallogr. Sect. A* **64**, 112-122 (2008).
49. Sheldrick, G. M. *SHELXL-97, Programs for X-ray Crystal Structure Refinement;* , University of Göttingen: Göttingen, Germany, (1997).
50. Barbour, L. J. *J. Supramol. Chem.* **1**, 189-191 (2001).
51. Dolomanov, O. V., Bourhis, L. J., Gildea, R. J., Howard, J. A. K. & Puschmann, H. OLEX2: a complete structure solution, refinement and analysis program. *J. Appl. Crystallogr.* **42**, 339-341 (2009).
52. Vignesh, K. R., Langley, S. K., Murray, K. S. & Rajaraman, G. Role of Diamagnetic Ions on the Mechanism of Magnetization Relaxation in "Butterfly" {Co$^{III}_2$Ln$^{III}_2$} (Ln = Dy, Tb, Ho) Complexes. *Inorg. Chem.* **56**, 2518-2532 (2017).
53. Singh, M. K., Yadav, N. & Rajaraman, G. Record high magnetic exchange and magnetization blockade in Ln$_2$@C$_{79}$N (Ln = Gd(iii) and Dy(iii)) molecules: a theoretical perspective. *Chem. Commun.* **51**, 17732-17735 (2015).
54. Singh, S. K., Gupta, T. & Rajaraman, G. Magnetic Anisotropy and Mechanism of Magnetic Relaxation in Er(III) Single-Ion Magnets. *Inorg. Chem.* **53**, 10835-10845 (2014).
55. Upadhyay, A. *et al.* Enhancing the effective energy barrier of a Dy(iii) SMM using a bridged diamagnetic Zn(ii) ion. *Chem. Commun.* **50**, 8838-8841 (2014).
56. Hess, B. A., Marian, C. M., Wahlgren, U. & Gropen, O. A mean-field spin-orbit method applicable to correlated wavefunctions. *Chem. Phys. Lett.* **251**, 365-371 (1996).
57. Roos, B. O. & Malmqvist, P.-A. Relativistic quantum chemistry: the multiconfigurational approach. *Phys. Chem. Chem. Phys.* **6**, 2919-2927 (2004).
58. Roos, B. O. *et al.* New Relativistic Atomic Natural Orbital Basis Sets for Lanthanide Atoms with Applications to the Ce Diatom and LuF$_3$. *J. Phys. Chem. A* **112**, 11431-11435 (2008).
59. Malmqvist, P. A., Roos, B. O. & Schimmelpfennig, B. The restricted active space (RAS) state interaction approach with spin-orbit coupling. *Chem. Phys. Lett.* **357**, 230-240 (2002).
60. Chibotaru, L. F. & Ungur, L. Ab initio calculation of anisotropic magnetic properties of complexes. I. Unique definition of pseudospin Hamiltonians and their derivation. *J. Chem. Phys.* **137**, 064112-064122 (2012).
61. Becke, A. D. Density functional thermochemistry. III. The role of exact exchange. *J. Chem. Phys.* **98**, 5648-5652 (1993).
62. Frisch, M. J. *et .al. Gaussian 09*, ***Rev. A.02***, (2009).
63. Dunning JR., T. H. & P.J. Hay. *In Methods of Electronic Structure Theory,* (H.F. SCHAEFER III, ED., PLENUM PRESS 1977).
64. Hay, P. J. & Wadt, W. R. Ab initio effective core potentials for molecular calculations. Potentials for the transition metal atoms Sc to Hg. *J. Chem. Phys.* **82**, 270-283 (1985).
65. Cundari, T. R. & Stevens, W. J. Effective core potential methods for the lanthanides. *J. Chem. Phys.* **98**, 5555-5565 (1993).
66. Noodleman, L. Valence bond description of antiferromagnetic coupling in transition metal dimers. *J. Am. Chem. Soc.* **74**, 5737-5743 (1981).
67. Blum, K. *Density Matrix Theory and Applications*. (Springer, 2012).
68. Rousochatzakis, I. & Luban, M. Master equations for pulsed magnetic fields: Application to magnetic molecules. *Phys. Rev. B* **72**, 134424 (2005).
69. Leuenberger, M. N. & Loss, D. Spin tunneling and phonon-assisted relaxation in Mn$_{12}$-acetate. *Phys. Rev. B* **61**, 1286-1302 (2000).
70. Lascialfari, A., Jang, Z. H., Borsa, F., Carretta, P. & Gatteschi, D. Thermal Fluctuations in the Magnetic Ground State of the Molecular Cluster Mn$_{12}$O$_{12}$ Acetate from mSR and Proton NMR Relaxation. *Phys. Rev. Lett.* **81**, 3773-3776 (1998).

## Acknowledgements

G.R. would like to acknowledge the financial support from DST-India, and IIT Bombay for the high performance computing facility. K.S.M and G.R. thank the Australia-India AISRF program for support. K.R.V. is thankful to the IITB-Monash Research Academy for a PhD studentship. G.R. acknowledges funding from SERB-DST (EMR/2014/000247) for financial support. AS acknowledges support from the Australian Research Council, Discovery Grant ID: DP15010325.

## Author contributions

A.S, K.S.M, S.K.L. and G.R. visualized and designed the project. K.R.V. and S.K.L carried out the syntheses and characterized the materials. S.K.L and K.R.V. performed the synchrotron X-ray scattering measurements, analyzed the data, and solved the crystal structure. W.W carried out micro-SQUID measurements. K.R.V. and A.S. carried out the *ab initio* calculations. A.S developed the theoretical models for magnetic coupling, and for the hysteretic dynamics of the magnetization, and used them for the simulation and interpretation of the magnetic data. A.S., K.S.M. and G.R. wrote the manuscript. All the authors discussed the results and contributed to the manuscript.

## Additional Information

**Supplementary Information:** http://www.nature.com/naturecommunications Structural (bond and angle data and packing diagrams), magnetic data (magnetization versus field, ac susceptibility, μ-SQUID and EPR measurements) and computational details (*ab initio* calculation of the magnetic properties of individual Dy sites; calculated low-lying spectrum of magnetic states of $\{Cr^{III}Dy^{III}_6\}$).

**Competing financial interests:** Authors declare no competing financial interests

**Reprints and permission** information is available online at http://npg.nature.com/reprints and permissions/
**How to cite this article:** Vignesh, K. R. et al. Ferrotoroidic Ground State in a Heterometallic $\{Cr^{III}Dy^{III}_6\}$ Complex Displaying Slow Magnetic Relaxation. *Nat. Commun.* X:XXXX doi: 10.1038/ncommsXXXX (2017).

## FOR TABLE OF CONTENTS ENTRY ONLY

**Heptanuclear {Cr-Ln$_6$} SMT:**

The first 3d-4f Single Molecule Ferrotoroidic (SMT) system has been synthesised, magnetic and theoretical studies have been thoroughly investigated.

**FOR TABLE OF CONTENTS ENTRY ONLY**

**Graphic**

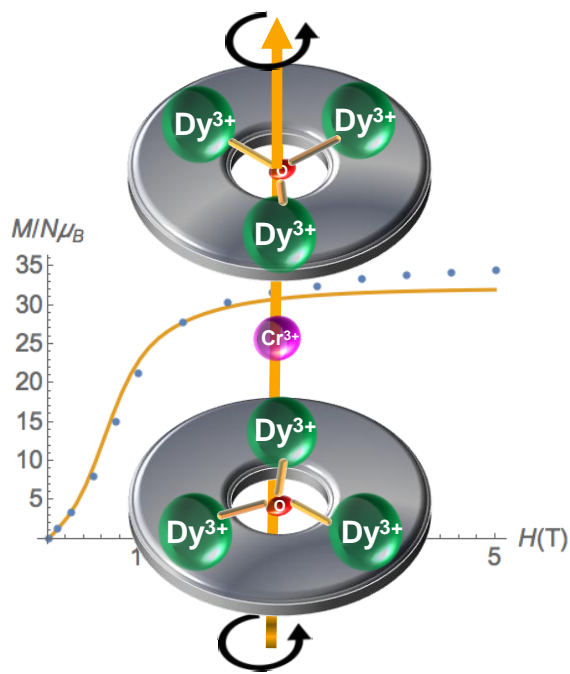